\documentclass[11pt, a4paper]{article} 
\usepackage[includeheadfoot,
            marginratio={1:1,2:3}, 
            width=450pt, 
            height=688pt,]{geometry}
\usepackage[skip=14pt plus2pt, indent=12pt]{parskip}

\usepackage{amsmath}
\usepackage{amsfonts}
\usepackage{amssymb}
\usepackage{bbm,bm}
\usepackage{mathtools}
\usepackage{slashed}
\usepackage{mathalfa}
\usepackage{graphicx,caption,subfigure}
\usepackage{tikz} 
\usepackage{tikz-cd}
\usetikzlibrary{decorations.pathreplacing,calc,tikzmark}
\usepackage{booktabs}
\usepackage{adjustbox}
\usepackage[utf8]{inputenc}
 \usepackage[splitrule]{footmisc} 
 
 \usepackage{comment}
 \usepackage{mathtools}
 \usepackage{tikz}
\usepgflibrary{arrows}
 
\usepackage{placeins}
\usepackage{empheq}
\usepackage{paralist}
\usepackage{cite}
\usepackage[normalem]{ulem}
\usepackage{soul}

\usepackage{xcolor}
\usepackage{tensor}
\usepackage{nicefrac}
\usepackage[colorlinks=false,urlbordercolor=purple
]{hyperref}

\newcommand{\nc}{\newcommand}
\nc{\lb}{\llbracket}
\nc{\rb}{\rrbracket}
\nc{\gl}{\llbracket}
\nc{\gr}{\rrbracket}
\nc{\del}{\partial}
\nc{\tri}{\hspace{-3.5pt}\vartriangle\hspace{-3.5pt}}
\nc{\blacktri}{\blacktriangle}
\nc{\ov}{\overline}
\nc{\fa}{\hat}
\nc{\fb}{\MakeUppercase}
\nc{\fc}{\tilde }
\nc{\Lie}{{\cal L}} 
\nc{\lambdabar}{{\mkern0.75mu\mathchar '26\mkern -9.75mu\lambda}}
\nc{\bh}{Schwarzschild-AdS$_{d}$ black holes }

\newenvironment{eqn}{\begin{equation}\begin{aligned}}{\end{aligned}\end{equation}}
\newenvironment{eqn*}{\begin{equation*}\begin{aligned}}{\end{aligned}\end{equation*}\noindent}

\allowdisplaybreaks[2]
\numberwithin{equation}{section}

\DeclareUnicodeCharacter{2212}{-}

\begin{document}
\vspace*{-1.5cm}
\begin{flushright}
  {\small
  MPP-2025-46\\
  LMU-ASC 07/25\\
  }
\end{flushright}

\vspace{1.5cm}
\begin{center}
  {\Large \bf Black Hole Transitions, AdS and the Distance Conjecture }
\vspace{0.35cm}
\end{center}

\vspace{0.35cm}
\begin{center}
{\large 
Alvaro Herráez$^a$, Dieter Lüst$^{a,b}$, and Carmine Montella$^a$
}
\end{center}

\vspace{0.1cm}

\begin{center} 
\emph{
$^a$Max-Planck-Institut für Physik (Werner-Heisenberg-Institut),
Boltzmannstr. 8, 85748 Garching, Germany,   \\[0.1cm]
 \vspace{0.3cm}
$^b${\it Arnold Sommerfeld Center for Theoretical Physics,\\
Ludwig-Maximilians-Universit\"at M\"unchen, 80333 M\"unchen, Germany}.}
\end{center}

\vspace{1.0cm}
\begin{abstract}
In this work, we investigate the connection between black hole instabilities and Swampland constraints, presenting new insights into the AdS Distance Conjecture. By examining the scale at which  horizon instabilities of Schwarzschild-AdS$_d$ black holes take place---$\Lambda_{\rm BH}$---we uncover a universal scaling relation, $\Lambda_{\rm BH}\sim |\Lambda_{\rm AdS}|^{\alpha}$, with $\frac{1}{d}\leq \alpha\leq \frac{1}{2}$, linking the emergence of towers of states directly to instability scales as $\Lambda_{\rm AdS}\to 0$. This approach circumvents the explicit dependence on field-space distances,  offering a refined formulation of the AdS Distance Conjecture grounded in physical black hole scales. From a top-down perspective, we find that these instability scales correspond precisely to the Gregory–Laflamme and Horowitz–Polchinski transitions, as expected for the flat space limit, and consistently with our proposed bounds. Furthermore, revisiting explicit calculations in type IIB string theory on AdS$_5\times S^5$, we illustrate how higher-derivative corrections may alter these bounds, potentially extending their applicability towards the interior of moduli space. Using also general results about gravitational collapse in AdS, our analysis points towards a possible breakdown of the conjecture in $d>10$, suggesting an intriguing upper limit on the number of non-compact spacetime dimensions. Finally, we briefly discuss parallel considerations and implications for the dS case.
\end{abstract}

%----------------------------------------------------------------------%
%  Resetting of counters
%----------------------------------------------------------------------%
%\setcounter{page}{1}
\pagestyle{empty}
\setcounter{footnote}{0}
%---------------------------------------------------------------%
%  Paper begins
%---------------------------------------------------------------%
\newpage
\tableofcontents
\pagestyle{empty}
\newpage
\pagestyle{plain}
\setcounter{page}{1}
\section{Introduction}\label{Introduction}

Within the Swampland program \cite{Vafa:2005ui},\footnote{See \cite{Brennan:2017rbf, Palti:2019pca, vanBeest:2021lhn, Grana:2021zvf, Agmon:2022thq,VanRiet:2023pnx} for some reviews.} it has become clear that not every low-energy effective field theory (EFT) can be consistently embedded into a quantum gravity theory in the ultraviolet (UV). Identifying the common features of EFTs that originate from quantum gravity is key for establishing a connection between high-energy quantum gravitational theories and the physics observed in the low-energy regime. 
These universal properties are frequently encapsulated by conjectures, which may be tested and, in certain cases, rigorously formulated within well-defined frameworks such as string theory or using the AdS/CFT correspondence. These hypothesis are generally motivated both by top-down arguments---such as string compactifications,  world-sheet analysis or AdS/CFT constructions---and by bottom-up ones---like black hole physics or unitarity constraints in scattering amplitudes. 
One of the most important Swampland conjectures is the Distance Conjecture \cite{Ooguri:2006in}, which states that in any theory consistent with quantum gravity, as one explores parametrically large distances in moduli space, an infinite tower of states becomes asymptotically light in the following way (in Planck units):
\begin{eqn} \lim_{\Delta_{\phi} \to \infty }m_{\rm tow}(\phi) \, \sim \, e^{-\lambda \Delta_{\phi}}\, , \end{eqn}
where $\Delta_{\phi}$ is the geodesic distance in moduli space and $\lambda$ is an order-one constant. From the perspective of string theory, the emergence of an infinite tower of light states is indicative of dualities. Thus, as an infinite-distance limit is approached, the original EFT breaks down and must be replaced by a dual description, governed by weakly coupled degrees of freedom associated with the lightest states in the tower.\footnote{See \cite{vandeHeisteeg:2022btw, Cribiori:2022nke, Calderon-Infante:2023uhz, Castellano:2023aum, vandeHeisteeg:2023dlw, Bedroya:2024uva, Castellano:2024bna, Calderon-Infante:2025ldq, Castellano:2025ljk, Calderon-Infante:2025pls} for recent discussions on the breaking scales of gravitational EFTs from different perspectives, including the interplay with black holes.}
In this context, the intricate structure of string dualities is reflected in the Emergent String Conjecture \cite{Lee:2019wij}, which states that these towers of light states correspond to either Kaluza-Klein (KK) modes---signifying the growth of an internal manifold---or higher-spin excitations of an asymptotically tensionless critical string, in some duality frame.

The Anti-de Sitter Distance Conjecture \cite{Lust:2019zwm} extends the principles of the Distance Conjecture to the context of asymptotically Anti-de Sitter spacetimes. It proposes that, for any family of AdS vacua with cosmological constant $\Lambda_{\rm AdS}$, approaching the limit $\Lambda_{\rm AdS} \rightarrow 0$ is inevitably accompanied by the appearance of an infinite tower of states, characterized by a mass scale $m_{\rm tow}$, such that (in Planck units) 
\begin{eqn} \label{ADC}
m_{\rm tow} \sim \left| \Lambda_{\rm AdS} \right|^\alpha \, , 
\end{eqn}
where $\alpha$ is an order-one constant. Initially, it was suggested that for supersymmetric vacua, $\alpha = \frac{1}{2}$, with the conjecture potentially extending to de Sitter spacetimes as well.  Successively, further studies based on the behaviour of the AdS Distance Conjecture under dimensional reduction pointed out a possible refinement $\alpha \geq \frac{1}{d}$ \cite{Rudelius:2021oaz,Gonzalo:2021fma,Castellano:2021mmx}. Substantial evidence has since supported both the Distance Conjecture and the AdS Distance Conjecture, based on string theory compactifications and through the AdS/CFT correspondence 
\cite{Blumenhagen:2017cxt, Grimm:2018ohb,Lee:2018urn, Lee:2018spm, Grimm:2018cpv, Buratti:2018xjt,Corvilain:2018lgw,Lee:2019tst, Joshi:2019nzi, Marchesano:2019ifh, Font:2019cxq, Lee:2019xtm, Baume:2019sry, Blumenhagen:2019vgj, Cecotti:2020rjq, Gendler:2020dfp, Lee:2020gvu,Lanza:2020qmt,Klaewer:2020lfg, Lanza:2021udy, Alvarez-Garcia:2021pxo, Palti:2021ubp,Etheredge:2022opl,Baume:2020dqd,Perlmutter:2020buo,Baume:2023msm,Calderon-Infante:2024oed}. 
Furthermore, applying the AdS Distance Conjecture to a positive cosmological constant, the smallness of the today's vacuum energy would imply the existence of a very light tower of states in our universe, an observation which has led to formulation of the
dark dimension scenario \cite{Montero:2022prj}.

Nonetheless, to get a broader understanding and avoid potential string lampost effects\footnote{See e.g. \cite{Montero:2020icj, Bedroya:2021fbu, Hamada:2021bbz, Cvetic:2020kuw} for recent discussions and evidence for \emph{string universality}, the idea that every consistent quantum gravity theory can be obtained from string theory.} it is essential to develop \emph{bottom-up} evidence for these conjectures. This approach has been pursued to find evidence for the Distance Conjecture from different perspectives \cite{Hamada:2021yxy,Stout:2021ubb,Stout:2022phm,Calderon-Infante:2023ler}, and in conjunction with gravitational entropy bounds for the Anti-de Sitter Distance Conjecture \cite{Castellano:2021mmx}. 

One of the main goals of our work is to provide an independent bottom-up argument for the Anti-de Sitter Distance Conjecture from a different prism---namely the physics of black holes, their instabilities and their internal consistency in an asymptotic AdS background---without relying on any specific UV completion. For this, we build on recent work  \cite{Bedroya:2024uva}, where it was highlighted that the appearance of a tower of states can manifest as an instability in semi-classical Schwarzschild black holes within the EFT framework. Specifically, in flat spacetime, the inverse length scale (or temperature) associated with the smallest black hole for which the EFT ceases to correctly capture the thermodynamic properties of the most stable configuration, denoted as $\Lambda_{\rm BH}$, indicates an instability or phase transition to a more stable configuration. In this context, the presence of a tower of states is linked to neutral black hole instabilities, a concept that, in principle, can be generalized across the moduli space. When the Emergent String Conjecture is applied to this scenario in the asymptotic flat-space limit (i.e., $\Lambda_{\rm AdS}\to 0$), it provides a refined understanding of $\Lambda_{\rm BH}$. In particular, this refinement identifies $\Lambda_{\rm BH}$ with the Gregory–Laflamme transition scale in the decompactification limits of quantum gravity, where $\Lambda_{\rm BH} \sim m_{\rm KK}$ asymptotically. On the other hand, in perturbative string limits, it connects $\Lambda_{\rm BH}$ to the Horowitz–Polchinski scale.  This yields $\Lambda_{\rm BH} \sim M_{\rm s} \leq T_{\rm Hag}$, where $M_{\rm s}$ is the string scale and $T_{\rm Hag}$ is the Hagedorn temperature---which is parametrically of the same order as $M_{\rm s}$, but strictly larger by a model-dependent order-one factor.

It is then natural to question whether this correspondence can be extended and generalized to other contexts, such as Anti-de Sitter spacetimes. In particular, one might ask whether, in the weak gravitational limit, the existence of neutral black hole instabilities can be constrained or linked to more general black hole instabilities or phase transitions in AdS. Following this reasoning, the AdS Distance Conjecture could be understood as a manifestation of the relationship between the black hole scale and the various black hole instabilities present in AdS. In short, by identifying the characteristic scale---$\Lambda_{\rm BH}$---from instabilities in neutral black hole solutions in AdS, and identifying it as a universal mass scale associated with a tower of states---$m_{\rm tow}$---in the limit $\Lambda_{\rm AdS} \rightarrow 0$, we naturally arrive at a refined version of the Anti-de Sitter Distance Conjecture, where $\frac{1}{d}\leq \alpha \leq \frac{1}{2}$. Moreover, no explicit reliance on string theory is required for the argument. 

More concretely, we argue that there exists both a lower and an upper bound on $\Lambda_{\rm BH}$, by analyzing the thermodynamic properties of Schwarzschild-AdS$_d$ black holes in thermal equilibrium with radiation, both in the canonical and in the microcanonical ensembles, following \cite{Hawking:1982dh,Witten:1998qj}. The lower bound corresponds to the critical temperature, $T_{\rm min}$---which represents the minimal temperature that a black hole can have in AdS--- and/or the Hawking-Page temperature, $T_{\rm HP}$ \cite{Hawking:1982dh}. As we argue in the following, the upper bound is instead given by the holographic constraint dictated by Jeans instability \cite{Jeans:1902fpv}, which in AdS is deeply related to the Breitenlohner–Freedman bound \cite{Breitenlohner:1982jf,Gross:1982cv}. Beyond this temperature, denoted $T_{\rm J}$, no meta-stable small black hole branch exists in AdS. Remarkably, from this interpretation of the AdS Distance Conjecture through the lens of black hole phase transitions and instabilities, we infer the allowed range $\frac{1}{d}\leq \alpha \leq \frac{1}{2}$, which is in strong agreement with string theory results.
This logic is similar to the one followed to provide recent bottom-up evidence for the Emergent String Conjecture. There, it has been studied that the tower of states can be understood explicitly through a Black Hole-Tower of States Correspondence \cite{Cribiori:2023ffn,Basile:2023blg,Basile:2024dqq,Herraez:2024kux}.\footnote{See also \cite{Bedroya:2024ubj,Kaufmann:2024gqo} for  bottom-up arguments for the Emergent String Conjecture from different approaches.}
In the latter,  a phase transition between minimal black holes and the tower of states at the species scale temperature, $\Lambda_{\rm sp}$, was proposed in the weakly-coupled gravitational limit. This  established a direct relationship between $ m_{\rm tow} \sim \Lambda_{\rm sp}^{\frac{d+p-2}{p}}$, with $p \geq 1\, , $ where $p$ parametrizes the number of large extra dimensions and also captures the emergent string limit as $p \rightarrow \infty$ \cite{Castellano:2021mmx}, from a bottom-up approach.

Furthermore, due to the UV/IR mixing inherent in these relations and their ties to quantum gravity phase transitions and instabilities, we discuss possible constraints on the number of non-compact dimensions for which the AdS Distance Conjecture holds. Along these lines, the conjecture seems not to  be valid for $d>10$. This is related to the transition of Einstein solutions from \emph{chaotic} behavior in lower-dimensional cases to non-chaotic behavior in higher dimensions \cite{Li:2009jq, Vaganov:2007at, Chavanis:2007kn ,Hammersley:2007ahw, Li:2008xw}, as well as the $d$-dependence of black hole-black string transitions \cite{Kol:2006vu, Sorkin:2004qq, Kol:2002xz, Harmark:2007md}.

Finally, we propose a potential generalization of the AdS Distance Conjecture via black holes, incorporating higher-derivative corrections, which extend the conjecture as one moves deeper into the moduli space.

Hence, this work is structured as follows: In section \ref{s:AdSBHchapter} we review the physics of Schwarzschild black holes and thermal configurations in $d$-dimensional AdS spacetimes, identifying key phase transitions and instabilities that arise due to the distinctive properties of this background. Next, in section \ref{thermoInstabilitiesChapter} we examine additional (thermo-)dynamic instabilities introduced by UV features motivated by Swampland principles, such as the presence of compact extra dimensions or light string modes. These are linked to the Gregory-Laflamme  and the Horowitz-Polchinski instabilities in AdS.
Finally, in section \ref{ADCandBHSection}, we present the derivation of the Anti-de Sitter Distance Conjecture in terms of black hole instabilities, and discuss a critical dimension beyond which this conjecture potentially breaks down. As part of this analysis, in section \ref{ss:higherderivative} we review previously computed higher-derivative corrections to $T_{\rm min}$ and $T_{\rm HP}$ in the AdS$_5 \times \text{S}^5$ setting. These modify the lower bound on $\Lambda_{\rm BH}$, which following our logic is the relevant scale to formulate the AdS Distance Conjecture. This allows us to conceptually extend the conjecture to the interior of moduli space. We conclude with some comments on the cosmological horizons and discuss possible generalizations in section \ref{ss:cosmologicalhorizons}, leaving some conclusions and future directions for section \ref{s:Summary}.

\section{Schwarzschild Black Holes in AdS}\label{s:AdSBHchapter}

In asymptotically flat space, a Schwarzschild black hole and thermal radiation can only be considered in the canonical ensemble at constant temperature if put inside a box \cite{Hawking:1976de}. This is due to the fact that the density of states of a Swcharzschild black hole in flat space grows super-exponentially, and thus overpowers the exponential Boltzman suppression, yielding a divergent canonical partition function unless cut-off by the introduction of a box---effectively introducing a maximum possible size for the black hole, and thus for its mass. Such configuration  in thermal equilibrium inside a box, although possible, turns out to be unstable due to the fact that as the mass of the black hole increases, its temperature decreases---i.e., it has negative specific heat, $\partial M/\partial T$.  Thus, a small fluctuation causing the black hole to radiate slightly less or more for a short period of time would cause the it to respectively grow or shrink indefinitely. If instead one considers the system inside a box in the microcanonical ensemble, and fixes the total energy of system consisting of the black hole plus the radiation, stable equilibrium between the two can be achieved provided the black hole has a significant portion of the total energy \cite{Hawking:1976de}. A natural objection to the consideration of such setup, though, is the fact that such a box may seem unphysical.  In AdS space, however, the thermal radiation remains confined\footnote{To be precise, non-zero rest mass particles are ``confined" in the AdS gravitational box, while zero rest mass particles can escape. However, in a thermal state the in-coming and out-coming fluxes at infinity are equal.} and is prevented from escaping to infinity due to the fact that the gravitational potential grows along the radial direction as $V(r) \sim +\, r^2/\ell_{\rm AdS}^2$. Roughly speaking, although the AdS volume is formally infinite, it provides an effective  gravitational box of radius  of  order $\ell_{\rm AdS}$, allowing for configurations in thermal equilibrium  at temperature $T$ \cite{Hawking:1982dh}.

The microcanonical and the canonical ensemble are therefore well-defined in asymptotically AdS spaces \cite{Hawking:1982dh,Witten:1998zw}. In  the former, it is interesting to study which are the most entropic phases depending on the total energy of the system, as they indicate the most stable saddle to which the other configurations are exponentially likely to tunnel.  In the later,  phase transitions at fixed temperature are indicative of a competition between various internal interactions within the system and play a crucial role in our understanding of its macroscopic and microscopic properties. This becomes particularly pertinent when universal relations and constants emerge from such investigations, such as critical exponents in mean field theory, which depend only on general features of the system but not on its particular details.
In the context of neutral black hole solutions and radiation in AdS, there are two parametrically-different, relevant  scales, which mark the transition between different regimes, both in the microcanonical and in the canonical ensemble, as originally described by Hawking and Page \cite{Hawking:1982dh}. In this section, we  review the relevant properties of thermal AdS and Schwarzschild-AdS black hole solutions in order to then study the scales at which thermodynamic instabilities signal transitions among them. We will mainly consider configurations at fixed temperature in the canonical ensemble, and sporadically comment on how similar results if the total energy is fixed instead.

\subsection{Thermal AdS$_d$ and Schwarzschild-AdS$_d$ black holes}
\label{s:ThermalAdSandAdSBHs}

Let us begin by recalling the thermodynamic properties of neutral black holes and radiation in asymptotically AdS spacetimes, highlighting the similarities and differences with respect to the asymptotically flat case,  mainly following \cite{Hawking:1976de,Hawking:1982dh,Witten:1998zw}. 

To begin with, we consider the metric of the covering space of AdS in $d$-dimensions with cosmological constant $\Lambda_{\rm AdS}$ in the static form, which reads \cite{Witten:1998zw}
\begin{eqn}
\label{eq:emptyAdSmetric}
    ds^2 &= - V(r) dt^2 + \frac{dr^2}{V(r)} + r^2 d\Omega_{d-2}^2, \\
    V(r) &= 1 + \frac{r^2}{\ell_{\rm AdS}^2}, \qquad \ell_{\rm AdS}^2 =- \frac{(d-2)(d-1)}{2\Lambda_{\text{AdS}}}\, ,
\end{eqn}
where $d\Omega_{d-2}$ is the metric on the $(d-2)$-dimensional sphere of unit radius, with area ${\Omega_{d-2}=2\pi^{\frac{d-1}{2}}/\Gamma\left(\frac{d-1}{2}\right)}$. The canonical ensemble is defined by an Euclidean path integral, with the Euclidean time compactified on a circle  of radius $\beta=T^{-1}$, over all matter fields and metrics which tend asymptotically to zero and AdS, respectively \cite{Hawking:1982dh}. The dominant contributions come from saddles that satisfy the Einstein equations with the aforementioned asymptomatics.  We follow here the convention of \cite{Hawking:1982dh}, in which the \emph{thermal AdS saddle}---coming from the metric \eqref{eq:emptyAdSmetric} upon compactification on the thermal circle---contributes to the euclidean partition function is considered as the zero of the free-energy, namely $F_{\rm AdS}=0$ for every $\beta$.\footnote{To be precise, this contribution is divergent due to the divergent volume of AdS, similarly to the ones from Swcharzschild-AdS black holes that we consider below. Since we are interested in differences between the corresponding euclidean actions, we can simply subtract the corresponding thermal AdS piece. The differences of Euclidean actions then turn out to be always finite for any $d>3$ \cite{Witten:1998zw}.}. This configuration prepares a thermal state in AdS with local temperature $T_{\rm loc}(r)\, =\, T\,  V(r)^{-1/2}$. Furthermore the corresponding entropy can be computed from the standard thermodynamic relations and yields $S_{\rm AdS}=0$. 

So far, we have considered the contribution from the saddle coming from the 
Euclidean AdS compactified on the circle, which corresponds to the tree-level contribution to the action, of order $(G_N)^{-1}$. One can then consider the one-loop fluctuations of the metric and matter fields around that saddle, which are expected to give a contribution of order $(G_N)^0$. For the sake of concreteness, we consider massless thermal radiation with negligible self-gravity on top of such AdS space. The metric \eqref{eq:emptyAdSmetric} is then modified as \cite{Hawking:1982dh, Page:1985em, Vaganov:2007at, Hammersley:2007ahw,Chavanis:2007kn,Li:2009jq}
\begin{eqn}
    V_{\mathrm{rad}}(r) = 1 + \frac{r^2}{\ell_{\rm AdS}^2} - \frac{G_N\, m_{ \mathrm{rad}}(r)}{ r^{d-3}},
\end{eqn}
where $G_N$ is Newton's constant\footnote{In this work we use the following normalization for the relation among Newton's constant, Planck length and Planck mass in $d$-dimensions: $8\pi G_N\,=\, \ell_{{\rm Pl},d}^{d-2}\, = \, M_{{\rm Pl},d}^{2-d}$} and $m_\mathrm{rad}(r)$ is the mass-distribution of the solution, such that $\lim_{r\rightarrow \infty} m(r)$ gives the total mass of the configuration, which must be finite. As a first approximation, one can consider the energy-momentum tensor with a new contribution given by the radiation pressure on AdS$_d$, namely ${p= \rho/(d-1)}\, $, yielding
\begin{eqn}
\label{eq:Tmunuradiation}
    T_\nu^\mu = \frac{\rho}{d-1}(\delta^\mu_\nu - d \delta_0^\mu \delta_\nu^0).
\end{eqn}
Here, $\rho (r) \simeq a_d \, T^d_{\rm loc} (r),$ is the local energy density of a thermal gas, with local temperature $T_{\rm loc} (r) = T\,  V_{\rm rad}(r)^{-1/2}$, and $a_d$ a $d$-dependent, order-one constant \cite{Vaganov:2007at}.
Up to $\mathcal{O}(1)$ constants, one can get a first estimate of the contribution of such configuration to the partition function by simply considering the thermal contribution without back reaction \cite{Page:1985em, Li:2008xw, Hammersley:2007ahw}. This gives an extra contribution to the free energy of the thermal ensemble on top of the thermal AdS saddle, which can be obtained from integrating the energy momentum tensor \eqref{eq:Tmunuradiation} over the thermal AdS background, yielding
\begin{eqn}
    |\log{Z_{\rm rad}}| \,=\, \mathcal{O}(1)\,   \Omega_{d-2}\, \int_{0}^\infty \,dr\, \frac{r^{d-2} }{\left(1+\frac{r^2}{\ell_{\rm AdS}^2} - \frac{G_N \, m_{\text{rad}}(r)}{r^{d-3}}\right)^{\frac{d}{2}}} \ \int_0^\beta d\tilde{\beta}\, {\beta}^{-d}  \, =\, \mathcal{O}(1) \left(\frac{\ell_{\rm AdS}}{\beta}\right)^{d-1}\, .
\end{eqn}
The first integral times the $\Omega_{d-2}$ factor comes from the metric contribution along the $(d-1)$ non-compact dimensions, whereas the second one is the thermal contribution from the euclidean $S^1$. The relevant thermodynamic quantities are then given by
\begin{eqn}
\label{eq:tempCanonical}
F_{\rm rad} \equiv &  -\dfrac{\log{Z_{\rm rad}}}{\beta}\,  =\, - \mathcal{O}(1) \dfrac{\ell_{\rm AdS}^{d-1}}{ \beta^{d}}\, , \\
     E_{\rm rad} \equiv & - \frac{\partial}{\partial\beta}\log Z_{\rm rad}  \, =\,  \mathcal{O}(1) \dfrac{\ell_{\rm AdS}^{d-1}}{\beta^{d}}\, ,  \\
    S_{\rm rad} \equiv & \, \beta   E_{\rm rad} + \log{Z_{\rm rad}}\, = \, \mathcal{O}(1)\left( \dfrac{\ell_{\rm AdS}}{ \beta}\right)^{d-1} \, .
\end{eqn}

In a similar manner, in a microcanonical ensemble with a large number of (micro)states in between $E$ and $E+d E$, the density of states $N(E)$ is given by the Laplace transform of the partition function, $ Z(\beta) = \int_0^\infty N(E) e^{-\beta E} dE\,$. For thermal radiation one obtains a saddle point for $N(E)$ whenever $\beta = \mathcal{O}(1)\, (l_{\rm AdS}^{d-1} E^{-1})^{\frac{1}{d}}\,$, which returns the same results as for the canonical ensemble calculation.

Let us now consider Schwarzschild-AdS$_{d}$ solutions. In global coordinates, their metric reads \cite{Hawking:1982dh, Witten:1998zw, Belhaj:2012bg} 
\begin{eqn} \label{metric}
    ds^2 =&  -f(r)dt^2 + \frac{dr^2}{f(r)} + r^2 d\Omega_{d-2} \,, \\
    f(r) =& 1 -  \frac{G_N M}{r^{d-3}} + \frac{r^2}{\ell_{\rm AdS}^2} \, ,\qquad \ell_{\rm AdS}^2 = -\frac{(d-1)(d-2)}{2\Lambda_{\text{AdS}}}\,,
\end{eqn}
where $M$ is the mass of the black hole.\footnote{In our convention  $M$ is related to the actual ADM mass of the black hole solution as $M = \frac{16\pi}{(d-2)\Omega_{d-2}} M_{\text{ADM}}$, with $\Omega_{d-2}$  defined below eq. \eqref{eq:emptyAdSmetric}. Notice that this means that the corresponding thermodynamic relations must be modified accordingly when written in terms of $M$. }
In Euclidean signature, this solution has a conical singularity at the horizon,  which is located at $f(r_h) = 0$, so that 
\begin{eqn}
    M = \dfrac{r_h^{d-3}}{ G_N}\left(1+\dfrac{r_h^{2}}{\ell_{\rm AdS}^2}\right)\, .
\end{eqn}
This singularity can be removed by identifying the (asymptotic) inverse temperature, $\beta$,  as 
\begin{eqn}\label{eq:temp}
		\beta = \frac{4\pi \; \ell_{\rm AdS}^2 r_h}{(d-1) \, r_h^2 + (d-3) \,\ell_{\rm AdS}^2}.
\end{eqn}
As oppossed to the thermal AdS case, in which the asymptotic temperature was a free parameter, here it is determined by the mass (or horizon radius) of the black hole. The  contribution from this saddle to the Euclidean partition function can be computed upon including the corresponding boundary terms \cite{PhysRevLett.28.1082, PhysRevD.15.2752}. Upon substracting the contribution from thermal AdS$_d$, the free energy and entropy of the Schwarzschild-AdS$_{d}$ solution in terms of $r_h$ and  $\beta$ reads \cite{Hawking:1982dh,Witten:1998zw}
	\begin{eqn}\label{eq:freeenerhy}
		F_{\rm BH} \, =\,   \frac{\Omega_{d-2} r_h^{d-3}}{16 \pi G_N}\left(1-\frac{r_h^2}{\ell_{\rm AdS}^2}\right)\,, \qquad
        S_{\rm BH}\, =\,   \frac{\Omega_{d-2} r_h^{d-2}}{4G_N}\,.
	\end{eqn}	

As opposed to the asymptotically flat case, \bh cannot exist at any temperature. Precisely from \eqref{eq:temp}, one can see that for each temperature two horizon radii are possible
\begin{eqn}\label{eq:rh(beta)}
    r_h(\beta) = \frac{2\pi \ell_{\rm AdS}^2}{(d-1)\beta}\left( 1 \pm \sqrt{1 - \frac{(d-3)(d-1)}{4\pi^2 \ell_{\rm AdS}^2}\beta^2}\, \right) \,,
\end{eqn}
and solutions exist only for $r_h>0$, that is, above a minimum temperature
\begin{eqn}
\label{eq:betamax}
    \beta < \dfrac{2\pi}{\sqrt{(d-1)(d-3)}}\  \ell_{\rm AdS} \equiv \beta_{\mathrm{max}}.
\end{eqn} 

One branch of solutions, given by the \emph{plus} component of the solution \eqref{eq:rh(beta)}, are the so-called \emph{large AdS black holes}. At high temperatures, $\beta \ll \ell_{\rm AdS}$, their horizon is $r_h \simeq \frac{4\pi \ell_{\rm AdS}^2}{(d-1)\beta}$. Their free energy and the entropy are (at leading order in $\beta/\ell_{\rm AdS}$)
	\begin{eqn}\label{eq:largeBH}
		F_\text{Large} \, =\,  & -\frac{1}{16\pi G_N} \left(\frac{4\pi}{d-1}\right)^{d-1} \ell_{\rm AdS}^{2(d-2)} \beta^{1-d}, \\ 
		S_\text{Large} \, = \, & \frac{1}{G_N} \left(\frac{4\pi}{d-1}\right)^{d-2} \ell_{\rm AdS}^{2(d-2)} \beta^{2-d}.
	\end{eqn}
For future use, let us note that the large AdS black hole solution is valid also around (and above) the Hagedorn temperature, as its horizon size there is huge. Surprisingly, its free energy is negative and its specific heat  positive, making it a stable solution for the system at high temperatures.
    
\begin{figure}[t!]
\begin{center}
       \includegraphics[width= 0.95\textwidth]{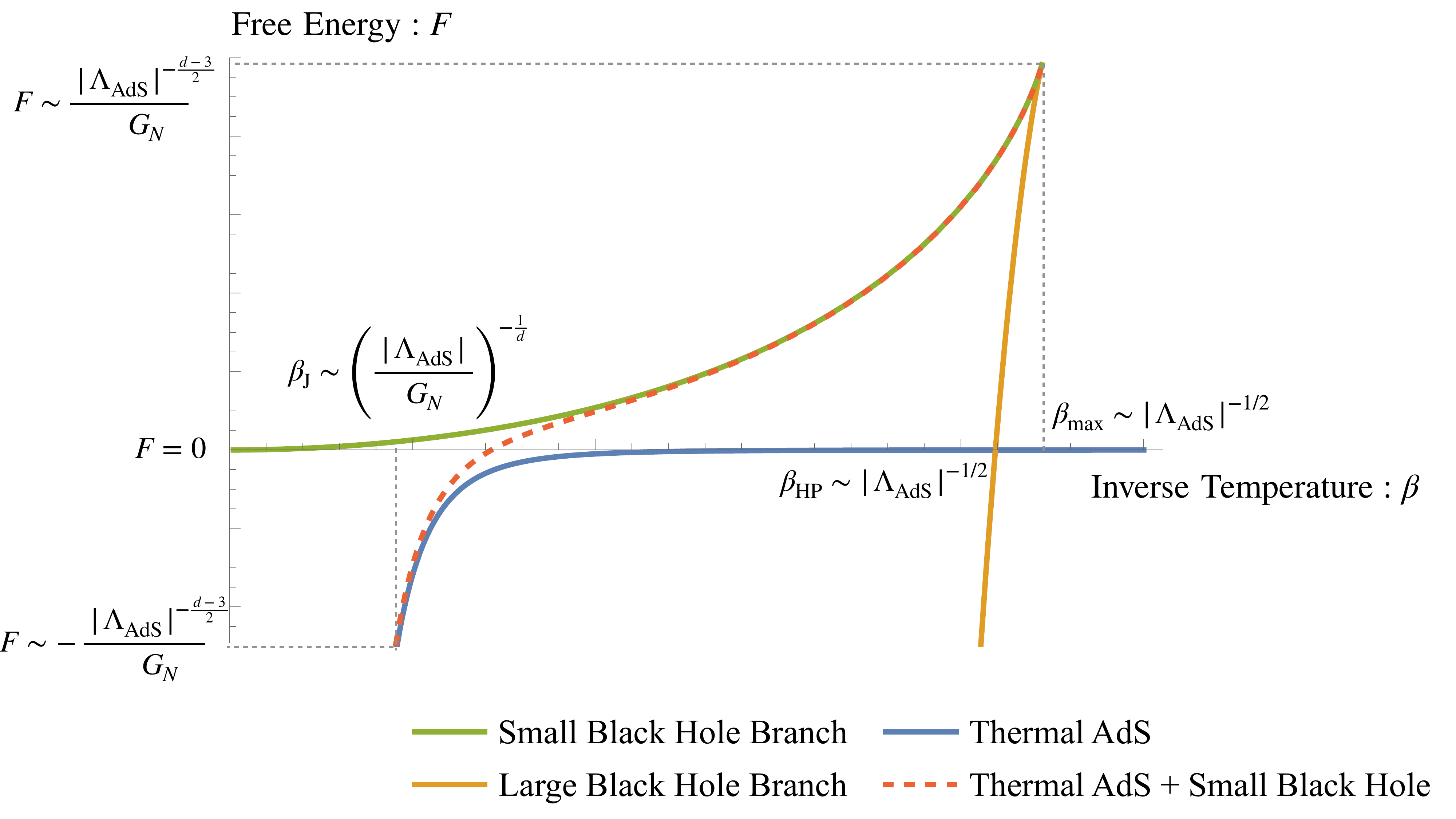}
	\caption{\small Schematic depiction of the relation between the free energy, $F$, and the inverse temperature, $\beta$, for the two neutral black hole branches and thermal radiation in AdS$_d$. For $\beta > \beta_{\rm max}$, no black hole solutions exist. For $\beta < \beta_{\rm HP}$, the large black hole branch has negative free energy and becomes the most favourable solution in the canonical ensemble. Around $\beta \sim \beta_{\rm J}$, the thermal AdS becomes unstable and collapses into the large black hole solution. } 
	\label{fig:F-beta_plot}
\end{center}
\end{figure}

The second branch, given by the \emph{minus} component in \eqref{eq:rh(beta)}, is usually dubbed \emph{small AdS black hole} solution. For temperatures $\beta \ll \ell_{\rm AdS}$ we find a horizon at $r_h = \frac{(d-3)}{4\pi} \beta$. As such, their free energy and entropy are given read (at leading order in $\beta/\ell_{\rm AdS}$)
	\begin{eqn}\label{eq:small_BH_S}
		F_\text{Small} \, = \, & \frac{1}{16\pi G_N} \left(\frac{d-3}{4\pi}\right)^{d-3} \beta^{d-3}, \\ 
		S_\text{Small}\,  = \, & \frac{1}{4 G_N} \left(\frac{d-3}{4\pi}\right)^{d-2} \beta^{d-2}.
	\end{eqn}
As one can see from the metric and the thermodynamic quantities, the small black hole branch behaves, at first order, just as a Schwarzschild black hole in asymptotically flat space time. A summary of the free-energy and as a function of the temperature for the relevant solutions reviewed here is presented in Figure \ref{fig:F-beta_plot}

\subsection{Transitions in Schwarzschild-AdS$_d$ Black Holes}\label{sec2}

The main goal of this work is to utilize the scales at which phase transitions among neutral black-holes and thermal AdS can take place to infer the scales capturing the limits for the regime  of validity of the underlying gravitational EFT, in the weak-coupling limit. To this end, we now review the main thermodynamic instabilities that affect such  AdS$_d$ solutions, highlighting the differences and similarities with respect to their asymptotically flat counterparts.

\subsubsection*{The Hawking-Page transition}
As one can see from eqns\eqref{eq:largeBH} and \eqref{eq:small_BH_S}, together with Figure \ref{fig:F-beta_plot}, the two black hole branches have very different thermodynamic properties. More precisely, \eqref{eq:largeBH} indicates that the large black hole solutions with $r_h \gg \ell_{\rm AdS}$ (i.e., with high temperature) have negative free energy, and positive specific heat. This implies that, in the canonical ensemble, they are a favourable configuration as compared to just thermal AdS$_d$. The temperature at which the free energy of the large Schwarzschild-AdS black hole solution equals that of thermal AdS is the \textit{Hawking-Page} temperature, $T_{\rm HP}$, which in $d$-dimensions takes the form \cite{Hawking:1982dh, Witten:1998zw}
\begin{eqn}
    T_{\rm HP} = \frac{d-2}{2\pi \, \ell_{\rm AdS}}\, ,  \qquad
    r_{h,\, {\rm HP}} = \ell_{\rm AdS}\, .
\end{eqn}
Note that the Hawking-Page temperature is only numerically larger than the minimal temperature for which Schwarzschild-AdS black hole solutions can exist, $T_{\mathrm{min}} = \beta^{-1}_{\rm max}\, $, namely
\begin{eqn}
    \frac{T_{\rm HP}}{T_{\mathrm{min}}} = \sqrt{\left(\frac{d-1}{d-2}\right)\left(\frac{d-3}{d-2}\right)}\, ,
\end{eqn}
which is greater than 1 for $d>3$. Remarkably, both temperatures have the same dependence on $\Lambda_{\rm AdS}$.
In the canonical ensemble, at temperatures  $0\leq T\leq T_{\rm min}$, thermal radiation is the only possible configurations. As the temperature of thermal AdS is raised to $T_{\rm min}\leq T\leq T_{\rm HP}$, black hole solutions (both for the large and small black hole branches) in thermal equilibrium with radiation are possible, but they are unstable, and the most likely configuration is still thermal AdS. As the temperature $T=T_{\rm HP}$ is reached, the thermal AdS undergoes a first-order phase transition, and for $T_{\rm HP}<T$ the most likely is that of a large black hole in thermal equilibrium with radiation. Hence, whereas $T_{\rm min}$ captures the minimum temperature for which both the neutral black hole solutions and the thermal AdS one exist, the Hawking-Page temperature encodes the minimum temperature for which the large black hole solutions are thermodynamically more favorable than thermal AdS$_d$.

\subsubsection*{The Jeans instability}
In AdS spacetimes, there are further potential instabilities that signal a limiting temperature for some of the phases in the theory. In particular, the thermal AdS$_d$ ensemble generically undergoes an instability caused by gravitational collapse under its own self-gravity \cite{Page:1985em,Vaganov:2007at,Hammersley:2007ahw,Chavanis:2007kn}, usually referred to as \emph{Jeans instability}. 
In order to understand when such instability may take place, one can estimate the gravitational backreaction of the finite energy density described by the energy-momentum tensor \ref{eq:Tmunuradiation} on the AdS background, which is a universal effect that must be taken into account every time one considers a thermal ensemble in the presence of gravity. As mentioned at the beginning of section \ref{s:ThermalAdSandAdSBHs}, thermal radiation in AdS is effectively confined to a region of radius of order $\ell_{\rm AdS}$. Thus, for a general thermal configuration, this instability can be characterized by considering the thermal AdS configuration whose total energy equals that of a black hole of radius $\sim \ell_{\rm AdS}$. This yields \cite{Misner:1973prb, Hawking:1982dh, Page:1985em, Vaganov:2007at,Hammersley:2007ahw,Chavanis:2007kn}
\begin{equation}\label{jeansMass}
   M_\mathrm{J} = \mathcal{O}(1)  \frac{\ell_{\rm AdS}^{d-3}}{G_N}\, ,
\end{equation}
such that thermal radiation can only exist for $E\lesssim M_{\rm J}$. When such critical mass is reached, the system collapses to a large black hole of AdS size due to its own self-gravity. This corresponds to a maximal temperature  for thermal radiation given by\footnote{For any equation of state $p = \omega \rho$ the exponent $\frac{1}{d}$ in \eqref{eq:maxTemp} is replaced by $\frac{\omega}{1+\omega}$. However, the ADM mass of such thermal configuration diverges for $\omega > \frac{1}{d-1}$, making $\omega=\frac{1}{d-1}$, which gives $\Lambda_{\rm AdS}^{1/d}$, the relevant upper bound, in agreement with our discussion. We elaborate on this in section \ref{ADCandBHSection}.} 
\begin{equation}\label{eq:maxTemp}
    T_{\rm J} = \mathcal{O}(1) \,  \dfrac{\ell_{\rm AdS}^{-\frac{2}{d}}}{G_N^{1/d}} \simeq \left(\dfrac{ |\Lambda_{\text{AdS}}|}{G_N}\right)^{\frac{1}{d}}\, ,
\end{equation}
This estimation was confirmed numerically in $d=4$ in \cite{Page:1985em}, and for general $d\geq 4$ in \cite{Vaganov:2007at,Hammersley:2007ahw,Chavanis:2007kn}. We will go back to the general $d$ case in section \ref{ss:MaxDim}. Note that even though $T_{\rm J}$ and $M_\mathrm{J}$ indicate the maximum mass and temperature at which thermal radiation can exist without collapsing into a black hole in AdS$_d$, both in the microcanonical (canonical) ensemble the thermodynamics are dominated by the black hole solutions---in equilibrium with thermal radiation---from energies (temperatures) well bellow these. Nevertheless, this limiting mass and temperature have a crucial physical meaning and play a key role in this work, as they differentiate the regime in which pure thermal radiation solutions can co-exist with black hole solutions---even if not-thermodynamically favoured---from that in which pure thermal radiation is no longer a solution.

Let us now briefly comment on a complementary way to obtain the same upper limit for the temperature---as discussed in \cite{Kleban:2013wba, Gribosky:1988yk, Kikuchi:1984np}. The key point is to realize that the negative curvature of AdS space affects stability with respect to its flat-space counterpart. Hence, scalars with negative mass may remain stable, provided their squared mass is above the Breitenlohner-Freedman (BF) bound in $d$ dimensions \cite{Breitenlohner:1982jf}:
\begin{eqn}\label{BFbound} 
m^2 \geq - \frac{d^2}{4\ell_{\rm AdS}^2}\, . 
\end{eqn}
As pointed out in \cite{Gross:1982cv, Gribosky:1988yk, Kikuchi:1984np}, in the local thermal bath, the component $g_{00}$ of the graviton acquires a tachyonic mass term at one-loop:
\begin{eqn}\label{tachmass} 
\delta m^2 = -\mathcal{O}(1) T^d\, , 
\end{eqn}
where the $\mathcal{O}(1)$ prefactor depends on the particular background under consideration. Consequently, Hawking radiation from small black holes in AdS at temperature $T$ receives a quantum correction from \eqref{tachmass}. Requiring then that the one-loop mas be above the BF bound implies
\begin{eqn}\label{maxtemp} 
T \leq T_{J} \sim \left|\Lambda_{\rm AdS}\right|^{\frac{1}{d}}\, . 
\end{eqn}
This represents a one-loop holographic bound on the surface located at radius $R_{\rm J}$, indicating that we cannot further increase the energy without triggering gravitational backreaction leading to collapse. If $T \gtrsim  T_{J}$, the radiation will collapse within a timescale  faster than $\ell_{\rm AdS}$, leading to the formation of a large black hole \cite{Barbon:2001di}.

Let us also mention that the same maximum mass, \eqref{jeansMass}, can be obtained by considering instead a system with a small Schwarzschild AdS$_d$ black hole (i.e. on with $r_{\rm h} \ll \ell_{\rm AdS }$), in local equilibrium with thermal AdS$_d$, instead of only thermal AdS$_d$. As mentioned in section \ref{s:ThermalAdSandAdSBHs}, this configuration is unstable and does not dominate the thermodynamics neither in the canonical nor in the microcanonical ensemble, but its collapse to a large black hole can nevertheless be studied. In the small $\Lambda_{\rm AdS}$ limit, $\ell_{\rm AdS} \gg \ell_{\rm Pl}$, the thermal radiation gives the dominant contribution to the ADM mass of the system. This is the case because the small black hole mass scales with the temperature as $M_{\rm Small}\sim \beta^{d-3}$, whereas the energy of the thermal AdS background goes as $E_{\text{AdS}}\sim\ell_{\rm AdS}^{d-1} \, \beta^{-d}$. Thus, the small black hole contribution to the total energy amounts to an additive constant, which is independent of $\Lambda_{\rm AdS}$\cite{Hawking:1982dh}, and this yields 
\begin{eqn}
    E = M_{\text{Small}} + E_{\text{AdS}} \simeq E_{\text{AdS}} \sim\ell_{\rm AdS}^{d-1} \, \beta^{-d} \, \quad \text{for} \quad \ell_{\rm AdS}\gg r_h\gtrsim \ell_{\rm Pl}\, .
\end{eqn}
One can then run the same argument as above to obtain the maximum temperature \eqref{eq:maxTemp} from equating the energy of the configuration to that of a black hole of size $\sim \ell_{\rm AdS}$.

\subsubsection*{Temperature range for co-existence of thermal AdS and black holes}
We have thus seen that, for general AdS$_d$ backgrounds, only thermal AdS can exist for $T\leq T_{\rm min}\sim |\Lambda_{\rm AdS}|^{1/2}$ (in Planck units)---since no black hole solutions exist for smaller temperatures, and only configurations with a large black hole can exist for $T\geq T_{\rm J} \sim |\Lambda_{\rm AdS}|^{1/d}$ (in Planck units)---as thermal radiation automatically collapses for higher temperatures.

In the canonical ensemble, the thermodynamics are dominated by the configuration with the lowest free-energy. In the region where both solutions can co-exist, this turns out to be the large black hole for $T_{\rm min}\simeq  T_{\rm HP} \leq T$, and thermal AdS for $T_{\rm min}\leq T\leq T_{\rm HP}$, as displayed in Figure \ref{fig:F-beta_plot}. Still, both the non-dominant of these two contributions and the (never dominant) small black hole configuration can exist in the aforementioned range of temperatures as unstable configurations in local thermal equilibrium, that can then decay to the one with lowest free-energy. 

In the microcanonical ensemble, the specifics differ slightly; however, the same range of temperatures is observed. In particular, for total energies of the order of $M_{\rm J}\simeq \ell_{\rm AdS}^{d-3}/G_N$, at which radiation collapses to a large black hole, one can find the maximum range of temperatures for the black hole and thermal AdS. The former has a temperature of the order $T\simeq T_{\rm min}\sim |\Lambda_{\rm AdS}|^{1/2}$ (in Planck units), and the latter $T\simeq  T_{\rm J}\sim |\Lambda_{\rm AdS}|^{1/d} $ (in Planck units). At lower energies, and in the region where both solutions exist, there is a regime where pure thermal radiation is more favorable---i.e., more entropic, and another one where the black hole in equilibrium with thermal radiation dominates the thermodynamics. Nevertheless, the temperatures of allowed configurations stay within the range $T_{\rm min }\leq T \leq T_{\rm J}$, or they to small for the black hole to exist. On the other hand, at energies above $M_{\rm J}$, no thermal AdS is possible, so only black hole configurations appear.

Even though one could be tempted to focus on the most favorable solutions alone, the less favourable ones still contribute to tunnelling amplitudes as $\Gamma \sim e^{-\Delta I}$, where $\Delta I$ is the difference between the actions of the two solutions. Thus, we have a range of temperatures defined by the possible existence of dynamical instabilities among black holes and thermal AdS, which we will link to the existence of an infinite tower of states with mass scale in the same range in section \ref{ADCandBHSection}.

\section{Towers of States and Black Hole Transitions in AdS$_d \times \mathcal{Y}^p$}
\label{thermoInstabilitiesChapter}
Analyzing the stability of a given spacetime is key to understanding multiple physical processes. In general relativity, stability indicates physical relevance, in the sense that a stable spacetime can arise dynamically from a wide variety of initial conditions. This issue becomes even more significant in the context of AdS/CFT, where the gravitational bulk can be described by the dynamics of gauge theories in the boundary of AdS of spacetime. In such scenarios, a gravitational instability in the bulk is linked to interesting gauge theory dynamics in the boundary, such as phase transitions. 

For the goal of this work we focus on dynamical instabilities of black hole spacetimes in AdS due to the presence of extra-dimensions---such as the Gregory-Laflamme instability \cite{Gregory:1993vy, Gregory:1994bj}---or weakly coupled string limits---such as the Horowitz-Polchinski transition \cite{Horowitz:1996nw,Horowitz:1997jc}. 
We also elaborate on the connection between these dynamical instabilities and the thermodynamical ones analyzed in section \ref{sec2}.

\subsection{Gregory-Laflamme Transition in AdS}

In asymptotically flat space time, the instability of non-extremal black branes with non-compact translational symmetry along the horizon was first explored in \cite{Gregory:1993vy, Gregory:1994bj}, providing strong evidence that long-wavelength perturbations destabilize these objects.\footnote{In this work, the end-point of the instability is not of fundamental relevance, so we will not elaborate on the recent results from \cite{Chu:2024ggi, Emparan:2024mbp,Chu:2025fko}. The relevant information for us is the scale at which the instability is triggered.} Roughly speaking, the instability arises when gravitational attraction cannot balance the object's tension along its extended direction, leading to an instability due to small perturbations of the horizon. 
In the presence of solutions wrapping compact dimensions, this instability suggests their non-viability within the low-dimensional EFT perspective, and the absolute necessity to consider a full higher-dimensional theory (see e.g. \cite{Emparan:2008eg} for an overview on black holes in higher dimensions and \cite{Calderon-Infante:2025ldq, Castellano:2025ljk} for recent discussions on the relevant scales for the matching of  black hole solutions in EFTs ). 

Such black brane solutions were then conjectured by Gubser and Mitra to become dynamically unstable under linear perturbations if and only if they happen to be (locally) thermodynamically unstable \cite{Gubser:2000ec, Gubser:2000mm}. Indeed, this connection can be made precise in AdS/CFT, since for spacetimes that are holographically dual to field theories, this manifests as an instability in the thermal ensemble of the corresponding field theory. More precisely, thermodynamic stability is equivalent to the requirement that the Hessian matrix of the entropy with respect to other extensive thermodynamic variables, such as mass and charge, $ \mathcal{H} = \left(\partial^2 S/\partial X^2\right)_{X = \{M,Q\}} \, $, must be negative definite \cite{Gubser:1998jb,Gubser:2000ec, Gubser:2000mm,Reall:2001ag}. This can be intuitively understood as the requirement that splitting the system into several parts, while conserving energy and charge, does not allow for an increase in entropy. Similar arguments, linking thermodynamic and dynamical instabilities, have been applied for non-extremal black brane inside a spherical cavity, where the crucial point is the key role that the boundary plays in affecting classical stability \cite{Gregory:2000gf}. Although the Gregory-Laflamme instability is typically linked to the dynamics on the horizon, different boundary conditions impose restrictions on allowed initial perturbations, similarly to the way in which internal space geometries play a role in restricting the perturbations that can arise in a system where they are present. 

In this sense, studying black holes or black branes in a box reveals several similarities to analyzing these same objects in asymptotically AdS spacetimes, where the \emph{AdS boundary} effectively plays the role of the \emph{box}.\footnote{Note that black holes in AdS have in general more complex thermodynamics, and phase transitions are also susceptible to corrections to the AdS background itself, as discussed in section \ref{alphaprimecorr} below.} Indeed, the Gregory-Laflamme instability has been explicitly studied in the context of asymptotically AdS space times \cite{Gregory:2000gf}. 
However, in AdS, for the large black hole branch the specific heat turns to be positive for $r_h\gg \ell_{\rm AdS}$, indicating stability under small perturbations according to the Gubser-Misra conjecture \cite{Gubser:2000ec, Gubser:2000mm}, and later confirmed by \cite{Reall:2001ag}. For this reason, we will focus on black holes with negative specific heat, i.e. black holes with radius $r_h\lesssim \ell_{\rm AdS} $, which are indeed not local minima of the Euclidean action.
The refinement and (semi-complete)proof of the conjecture was presented in \cite{Reall:2001ag} by demonstrating the equivalence between the classical Lorentzian \emph{threshold unstable mode}---which signals the dynamical Gregory-Laflamme instability---and the \emph{Euclidean negative mode}---related to thermodynamic instability \cite{Reall:2001ag, Gregory:2011kh}, as we briefly review in the following.

\subsubsection*{Relating dynamical and thermodynamic instabilities with extra dimensions}
Let us consider a general spacetime (in the presence of extra dimensions), ${\bf B}_d\times \mathcal{Y}^p$, where ${\bf B}_d$ represents a $d$-dimensional black hole spacetime---whose coordinates we label by $x$--- whereas $\mathcal{Y}^p$ stands for a $p$-dimensional transverse space---with coordinates $y$.
Then, the critical mass for the $(d+p)$-dimensional black hole spacetime is defined such that a perturbation mode, $h_{\mu\nu}(x, y)$,  marginally becomes tachyonic under the Euclidean Lichnerowicz operator \cite{Hubeny:2002xn, Kol:2004pn, Gibbons:2002pq}
\begin{eqn*}
    \Delta_L = \delta^{a}_{m} \delta^{b}_{n} \Delta_{L; ab}^{mn} = -\delta^{a}_{m} \delta^{b}_{n} \nabla^2 -\delta^{a}_{m} \delta^{b}_{n} R_{ab}^{\;\;\; mn}
\end{eqn*}
where the first term represents the free propagator and the second includes the contribution from the non-vanishing curvature. Hence, the marginal perturbation is defined by
\begin{eqn}\label{tachMode}
    \Delta_{L}h_{\mu\nu}(x, y) = 0\, .
\end{eqn}
Due to the product structure of spacetime, we can decompose $h_{\mu\nu}(x,z) = h_{\mu\nu}(x) \phi(y)$, and the Euclidean Lichnerowicz operator acts as
\begin{equation}
    \Delta_L h_{\mu\nu}(x,y) = \phi(y) \Delta_{L} h_{\mu\nu}(x) +  h_{\mu\nu}  \Delta_{L} \phi(y).
\end{equation}
For simplicity, we consider here a Ricci-flat $\mathcal{Y}^p$, which implies $\Delta_L \phi = \nabla^2 \phi$.
such that the Lichnerowicz operator on ${\bf B}_m$ and $\mathcal{Y}^p$ act non-trivially as
\begin{equation}
    \Delta_L({\bf B} )\  h_{\mu \nu} \, =\,  \lambda_E \, h_{\mu \nu}\, ,  \qquad  
 \nabla^2(\mathcal{Y})\  \phi \, =\,  - \lambda_X \, \phi\, .
 \end{equation}
For \eqref{tachMode} to be satisfied the two eigenvalues must be equal, and the instability then emerges when the Euclidean eigenmode is saturated by the momentum in the $X^n$ directions, i.e., $\lambda_E = \lambda_X$.

As shown in \cite{Reall:2001ag, Gregory:1993vy}, for an uncharged black $p$-brane with ansatz $h_{\mu\nu}(x,y) = e^{i y_k \mu_k} h_{\mu\nu}(x)$, the Lorentzian Lichnerowicz operator eigenvalue equation reads
\begin{equation}
\tilde{\Delta}_L h_{\mu\nu}(x) = -\mu^2 h_{\mu\nu}(x)\, .
\end{equation}
Crucially, there is a critical value $\mu^*$ such that for $\mu < |\mu^*|$, the eigenvectors grow exponentially in time, i.e.,  $h_{\mu\nu}(x) \propto e^{\Omega t}$, indicating an instability. Due to the symmetry of the solution, the Wick rotation just changes the sign of $h_{tt}$, which implies that $\lambda_E = - \mu^2$, and the Lorentzian unstable mode thus corresponds to an Euclidean negative mode.

Let us consider, for concreteness, the example of a black hole in $d$-dimensional asymptotically flat spacetime with $p$ extra dimensions compactified on a torus, $\mathcal{T}^p$, such that $\phi(\vec{y}) = \phi(\vec{y} + \vec{e})$. In this case it is possible to show that $\lambda_E = k_d/r_h^2$, with  $k_d \simeq \sqrt{d-3}$ an $\mathcal{O}(1)$ constant independent of the Schwarzschild radius,  $r_h$ \cite{Kol:2004pn}. Thus, there is a value of the Schwarzschild radius below which the black hole solution becomes unstable, obtained from the minimum non-zero eigenvalue of $\nabla^2(\mathcal{T}^p)$ as
\begin{equation}
    r_{h,\, \mathrm{crit}} = \left( \frac{k_d}{\min{\lambda_{\mathcal{T}^p}}}\right)^{\frac{1}{2}}\, .
\end{equation}
As explained above, in Lorentzian signature, the eigenvalues of the Lichnerowicz operator are related to the Euclidean ones as $\lambda_E = -\mu^2$. In this setup, the critical value $\mu_{\rm crit}$ is such that, for $\mu < |\mu_{\rm crit} |$, the eigenvectors of the Lorentzian Lichnerowicz operator grow exponentially in time, indicating an instability. This occurs when 
\begin{equation} 
|\mu_{\rm crit} | = \left(\dfrac{\min{\lambda_{\mathcal{T}^p}}}{k_d}\right)^\frac{1}{2}  \, ,
\end{equation} 
which gives the minimum Kaluza-Klein scale. As a consistency check, we can consider the limit in which the latter is very small, which results in the absence of an instability, due to $\mu_{\rm crit} \rightarrow 0$.

In general, this argument can be generalized for more general set-ups to include any compact manifold. For instance, for non-flat $\mathcal{Y}^p$ one should consider the Ricci part of the Lichnerowicz operator for $\phi(y)$, which will modify the spectrum $\lambda_X$.\footnote{In the large volume limit, this extra contribution will approximately depend on the characteristic length $R$ of the manifold as $\lambda_X \simeq \lambda_X^{\text{flat}} + {\mathcal{O}(1/{R^2})}$.}

\subsubsection*{The Gregory-Laflamme Instability and the Jeans Instability in higher dimensions}
\label{sss:GLandJeans}
Let us  begin by considering the example of AdS$_5 \times S^5$. This is one of the best studied AdS backgrounds in the context of AdS/CFT,  and since $R_{S^5}\simeq \ell_{\rm AdS}$ it does not present scale separation. For concreteness, we focus on black hole solutions with the horizon \emph{smeared} over the $S^5$. Such solutions lack translational symmetry, but still have a compact $SO(6)$ symmetry from the horizon being smeared along the compact dimensions. As shown in \cite{Hubeny:2002xn, Buchel:2005nt, Herzog:2017qwp}, in the AdS$_5 \times S^5$ set-up there is a static mode in Lorentzian geometry, marking the threshold between stable and unstable phases, with the first unstable mode appearing at around $0.426 \ell_{\rm AdS}$, which corresponds to the small black hole branch. It is then interesting to assess the stability of the small black hole branch for radii comprised between the maximum possible one in AdS$_5$ and the aforementioned threshold unstable mode, i.e. for  $ 0.426 \ell_{\rm AdS}\leq r\leq r_{h,\, {\rm max}}=\ell_{\rm AdS}/\sqrt{2}\simeq 0.707 \ell_{\rm AdS}$. The associated state in gauge theory continues to exhibit the full $SO(6)$ symmetry. For $r_+ = 0.426 \ell_{\rm AdS}$, however, a static threshold unstable mode emerges, which deforms the homogeneously smeared horizon along the $S^5$ into an inhomogeneous solution with the shape given by the first spherical harmonic. As $r_h$ decreases below $0.426\ell_{\rm AdS}$, dynamical instability leads to an exponentially growing mode. Incidentally, if boundary conditions at infinity are imposed to prevent the condensation of the $l=1$ mode, higher harmonics may condense instead, leading to a tower of phase transitions in the gauge theory for progressively smaller values of $r_h$.  Interestingly, \cite{Buchel:2015gxa} extended this analysis to generic compact manifolds $\mathcal{V}_5$, showing that the quasinormal mode equation governing the instability is universal, with the instability's onset determined by the lowest non-vanishing eigenvalue of the Laplacian. Similar reasoning can be applied to AdS$_{4,7} \times S^{7,4}$ as discussed in \cite{Horowitz:2000kx}.

In terms of the temperatures, while the black holes exists for $T \geq T_{\mathrm{min}} = \frac{\sqrt{2}}{\pi \ell_{\rm AdS}} = \frac{0.45}{\ell_{\rm AdS}}$, the small black hole branch only becomes Gregory-Laflamme unstable via localization on the $S^5$ for $T \geq T_{\rm GL}\simeq  \frac{0.53}{\ell_{\rm AdS}}$. Additionally, the Hawking-Page transition, involving the large hole branch, occurs at $T_{\rm HP} = \frac{0.48}{\ell_{\rm AdS}}$, such that we get $T_{\rm min}\leq T_{\rm HP}\leq T_{\rm GL}$, but all of them scale as $|\Lambda_{\rm AdS}|^{1/2}$.

The main lesson that we want to draw from this is the fact that the temperature at which the Gregory-Laflamme transition generally takes place can be bounded from above by the Jeans temperature in the corresponding higher-dimensional theory. In fact, the upper limit on the temperature coming from the Jeans instability in pure AdS$_d$, $T_\mathrm{J}\simeq |\Lambda_{\rm AdS}|^{1/d}$ (c.f.  eq. \eqref{eq:maxTemp}), presents a general upper bound for the temperature for collapse of thermal radiation in the presence of extra dimensions, which at the same time turns out to bound $T_{\rm GL}$ from above. The case ${T_{\rm GL}\sim |\Lambda_{\rm AdS}|^{1/2}}$ presented above, in which the internal radius is of the same order as that of AdS, turns out to be a limitting instance of the aforementioned general result. 

To illustrate this, let us consider a compactification on AdS$_d \times \mathcal{Y}^p$, with the internal volume given by $R^p$, with $R\leq \ell_{\rm AdS}$ its characteristic length scale, such that $T_{\rm GL}\simeq R^{-1}$.\footnote{For simplicity we restrict ourselves here to the homogeneous case, in which there is only one scale determining the size of the extra dimensions and thus signaling the onset of the Gregory-Laflamme instability. More general cases can be accommodated by labeling by $R$ the largest of these length scales and adapting the $p$ accordingly.} Thermal radiation in such background is effectively confined to a box of size $\ell_{\rm AdS}^{d-1}\times R^p$, and thus the Jeans mass of the black brane of similar size is 
\begin{equation}
    M_{\rm J}\simeq \dfrac{\ell_{\rm AdS}^{d-3}\,  R^p}{G_{N,\, d+p}}=\dfrac{\ell_{\rm AdS}^{d-3}}{G_{N,\, d}}\, ,
\end{equation}
where we have used the relation between the $(d+p)$- and the $d$-dimensional Newton constants, namely $G_{N,\,  d+p}\, =\,  G_{N,\,  d}\,  R^p$. This means that the energy at which the radiation collapses is the same as the one obtained in the $d$-dimensional theory, as the change in Planck units is precisely accounting for the homogeneous distribution of such radiation along the extra dimensions. That is, the Jeans mass for such distribution coincides exactly with that in \eqref{jeansMass}.

However, recall that the relation between the temperature and the energy of thermal radiation on a box of volume $\mathrm{Vol}_{D-1}$ in $D$ spacetime dimensions is given by $E\simeq \mathrm{Vol_{D-1}} \, T^{D}$, and thus explicitly depends on $D=d+p$. This implies that the corresponding Jeans temperatures will, in general, be different. In fact, the higher dimensional Jeans temperature---expressed in lower dimensional Planck units to facilitate comparison with \eqref{eq:maxTemp}---takes the form
\begin{equation}
    T_{\mathrm{ J},\, d+p }\simeq \left(\dfrac{\ell_{\rm AdS}^{-2} \ R^{-p}}{{G_{N,\, d}}}\right)^{\frac{1}{d+p}}\,.
\end{equation}
For the asymptotic regime we are interested in for this work, namely $\ell_{\rm AdS}\gg \ell_{\mathrm{Pl},d}$, this is smaller than $T_{\mathrm{J},\, d}$ for all $R\geq \ell_{\rm AdS}^{2/d}$. This means that, as the temperature of the thermal radiation is raised, it will gravitationally collapse at temperatures below $T_{\rm J,\, d}\simeq |\Lambda_{\rm AdS}|^{1/d}$ (in $d$-dimensional Planck units) if the inverse size of the extra dimensions is below that scale, i.e. $R^{-1}\lesssim T_{\mathrm{J},\, d}$. From the $d$-dimensional point of view, this can be seen as the fact that as the temperature reaches  $R^{-1}$, KK species will start to  contribute to the thermodynamics of the system, and thus the Jeans mass can be reached for lower temperatures \cite{Herraez:2024kux}. Precisely when the size of the compact dimensions is of order $R\simeq \ell_{\rm AdS}^{2/d}$, we have $T_{\mathrm{J},\, d}\simeq T_{\mathrm{J},\, d+p}\simeq T_{\rm GL}\simeq |\Lambda_{\rm AdS}|^{1/d}$ and the gravitational collapse will take place nevertheless at that temperature because the system will gravitationally collapse in $d$-dimensions. Following the logic that these limiting temperatures in which thermal radiation in AdS can coexist with black holes identify the possible ranges for the scale of the instabilities---that we will later identify with the towers---then gives the following bounds (in $d$-dimensional Planck units)
\begin{equation}
    T_{\mathrm{J},\, d}\, \simeq\,  |\Lambda_{\rm AdS}|^{1/d} \, \gtrsim \, T_{\mathrm{J},\, d+p}\,  \gtrsim \, T_{\rm GL}\, \simeq \, R^{-1}\, \gtrsim \,  T_{\rm HP}\, \simeq \, T_{\rm min }\, \simeq \, |\Lambda_{\rm AdS}|^{1/2}\, .
\end{equation}
To sum up, the upper bound for the Gregory-Laflamme temperature comes from gravitational collapse of thermal radiation in AdS$_d\times \mathcal{Y}^p$, which is at the same time upper bounded by that of pure AdS$_d$ and gives a minimum size for the extra dimensions. The lower bound comes instead from the existence of black hole solutions, and can be mapped to an upper bound for the size of extra dimensions, that must not exceed the characteristic length scale of the non-compact AdS$_d$. 

\subsection{Horowitz-Polchinski Transition in AdS}
In flat spacetime, fundamental long strings exhibit a vast degeneracy of states that grows exponentially with their mass. For a long, highly-excited, free string of length $L$, its mass and entropy can be analyzed via a random walk model. The total length and mass of such configurations can be understood by summing over $N$ steps, each corresponding to the string length scale $\ell_{\rm s}$, leading to an effective length $L \sim N \ell_{\rm s}$ and mass $M \sim 
\frac{N}{\ell_{\rm s}}$. The number of possible configurations grows exponentially with $N$, resulting in an entropy $S_s \sim N$ \cite{Susskind:1993ki, Susskind:1993ws,Horowitz:1996nw, Horowitz:1997jc}: \begin{eqn} S_s \sim \frac{M}{M_{\rm s}}. \end{eqn} This differs from black hole entropy, which scales as: \begin{eqn} S_{\rm BH} \sim \left( \frac{M}{M_{\rm pl, d}}\right)^{\frac{d-2}{d-3}}. \end{eqn}

Given the distinct entropy behaviours, it is common to hypothesize a critical mass, $M_{\mathrm{crit}}$, at which the entropy of a black hole and that of a string state become comparable. In other words, the Bekenstein-Hawking entropy bound may impose a cutoff, preventing the formation of non-black hole states with higher energies. Consequently, large black holes would stop their Hawking evaporation process at $M = M_{\mathrm{crit}}$, transitioning into a higher-entropy string state without ever reaching the singular zero-mass limit.

This ``correspondence principle" was conjectured in \cite{Horowitz:1996nw, Horowitz:1997jc, Horowitz:1999uv, Susskind:1993ki, Susskind:1993ws, Sen:1995in}, suggesting that the spectrum of black holes and that of string states are ``identical" in the sense that there is a one-to-one correspondence between fundamental string states and black hole states 

A canonical formulation of this principle is presented via the thermal string theory partition function on asymptotically $R^d \times S^1$. Near the Hagedorn temperature, where $\beta - \beta_H \ll \ell_{\rm s}$, the first string winding mode around the thermal circle becomes much lighter than the string scale. Upon compactifying the thermal circle, bound-state solutions of the winding scalar with gravity emerge \cite{Damour:1999aw, Khuri:1999ez, Kutasov:2005rr, Giveon:2005jv, Chen:2021emg, Brustein:2021cza, Chen:2021dsw}. These solutions describe self-gravitating bound states of hot strings, referred to as ``string stars."\footnote{It is worth noting that \cite{Chen:2021emg} observed that the correspondence principle works smoothly for heterotic strings, but does not seem to apply directly in the same way for type II strings. This discrepancy may be due to the involvement of D-branes in the transition.}

In the context of (Asymptotically-)Anti-de Sitter spacetime, unlike in asymptotically flat spacetime, the canonical partition function is well-defined. The radius of the thermal circle is fixed only at the conformal boundary of space, making the bulk partition function non-strictly thermal.
The Horowitz-Polchinski transition in AdS has been studied in \cite{Urbach:2022xzw} showing that at sufficiently low temperatures, where $\ell_{\rm s}^2 /\ell_{\rm AdS} \ll \beta - \beta_{\text{Hag}} \ll \ell_{\rm s}$, the sizes of both the black hole and string star solutions become very small and the two solutions coincide. At higher temperatures, the AdS string star expands with increasing temperature but is constrained by a maximum size of $L \leq L_c \sim \sqrt{\ell_{\rm s}\ell_{\rm AdS}}$, while for the flat-space string star grows until its length scale diverges. However, at a critical temperature above the flat-space Hagedorn temperature, $\beta_c < \beta_H$, which corresponds to the Hagedorn temperature in AdS space, the solution reaches its maximum size. At higher temperatures $\beta < \beta_c$ the string star’s amplitude goes to zero and the solution smoothly joins with the thermal gas phase\cite{Urbach:2022xzw}.
Additionally, in AdS, string star solutions are viable near $\beta_c$ for all $d$. For $d \geq 6$, the solution becomes unreliable as the temperature approaches the flat-space Hagedorn temperature $\beta_H$, and it ceases to exist at lower temperatures, representing a version of the correspondence principle specific to AdS space\footnote{See \cite{Bedroya:2024igb} for recent results on string stars for $d>6$ in flat-spacetime.}.
The connection between the maximal length of the string star in AdS nicely fits in the context of the AdS Distance Conjecture in terms of black holes. Indeed, due to the maximum size $L_c$, we can infer that $\ell_{\rm s} \leq L_c$, which  implies $\Lambda_{\text{AdS}}^{\frac{1}{2}} \lesssim M_{\rm s} $. This is indeed the requirement that the temperature of the string stars must remain above
$T_{\rm min}$.

However, other saddle points can affect the dynamics. For example localized black hole may appear first, or at the same scale, respect to the string-star saddle point, which make the system more complicated to study. In the limit $V_p^{\frac{1}{p}} \ll \ell_{\rm s}$, where $V_p$ is the volume of the $p$-compact extra dimensions, we would be sure that stringy effects have bigger contributions, and the previous analysis\cite{Urbach:2022xzw} would not change. 
However, String-star can suffer of Gregory-Laflamme, when a black string reaches string-scale thickness along a portion of its horizon. Then we can regard that part of the black string as having transitioned into an object made of highly excited string that extends along a linear direction\cite{Emparan:2024mbp}.

\section{Black Holes and the Anti-de Sitter Distance Conjecture}\label{ADCandBHSection}

In its original version \cite{Lust:2019zwm}, the Anti-de Sitter Distance Conjecture was motivated from a string theory perspective, and through the Distance Conjecture \cite{Ooguri:2006in}. The main result of this conjecture is that the characteristic mass of the lightest (infinite) tower of species is connected to the Anti-de Sitter cosmological constant, $\Lambda_{\rm AdS}$,  via
\begin{eqn}\label{AdSconjecture}
    m_{\rm tow} \sim \left|\Lambda_{\rm AdS}\right|^\alpha \, ,
\end{eqn}
in $d$-dimensional Planck units and in the limit in which $\Lambda_{\rm AdS}\ll M_{{\rm Pl, d}}^2$. 

Furthermore,  it was also argued that for supersymmetric vacua $\alpha = 1/2$. Further studies on the allowed values for $\alpha$ also proposed $\alpha \geq 1/d$, with $d$ the number of non-compact dimensions\cite{Rudelius:2021oaz, Gonzalo:2021fma,Castellano:2021mmx}. 
A key point about the AdS Distance Conjecture is that the dependence on the distance in the field space, $\Delta_\phi$, is implicit,\footnote{For recent proposals for a general notion of distance away from moduli spaces see \cite{Kehagias:2019akr,Li:2023gtt,Basile:2023rvm,Demulder:2023vlo,Palti:2024voy,Mohseni:2024njl, Debusschere:2024rmi,Demulder:2024glx}} and the relation between two energy scales is manifest. Hence it may be suitable for an understanding in terms of black hole scales.

In this section, we will denote the important black hole-related scales by $\Lambda_{\rm x}(\phi)$, rather than $T_{\rm x}$, as we did in previous sections. This is to emphasize that $\Lambda_{\rm x} (\phi)$ refers to the full quantum-corrected black hole scales, which are in general field dependent an in the weak gravitational regime reduce to the temperatures studied in the previous sections. In the context of string theory, this refers to formulating the AdS Distance Conjecture through black hole scales, which are field-dependent and, in principle, computable within a given EFT. We return to this point in section \ref{ss:higherderivative}.

\subsection{Black Hole Argument for the AdS Distance Conjecture}

Our proposal for a bottom-up understanding of the conjecture lies on the interpretation of the AdS Distance Conjecture as a relation between the scales of \textit{dynamical} instabilities of neutral black holes in AdS. The key point is the use of the idea that any instability in these neutral black holes that is not describable in the original EFT, indicates the presence of a new tower of degrees of freedom \cite{Bedroya:2024uva}. Thus, if any such instability is present, it must be comprised between the scales where these black holes can exist, namely between $\Lambda_{\rm min} \sim \Lambda_{\text{HP}}$, and the scale at which thermal AdS does not allow any meta-stable small black hole solution due to gravitational collapse, i.e., the Jeans temperature $\Lambda_{\text{J}}$.\footnote{In principle, those quantities are related to the EFT action, and they acquire higher-derivative correction at fixed cosmological constant, and in principle it may allow us to extended them closer to the bulk of the field space. We elaborate on this in an example in the next section.}\footnote{This is in agreement with complementary arguments to motivate the bound $\alpha\geq 1/d$ in \cite{Castellano:2021mmx},  which related the scale of gravitational collapse of particles in a box of size $\ell_{\rm AdS}$ to the mass of the lightest state in a tower.}

More concretely, as argued in section \ref{thermoInstabilitiesChapter}, the scale of the potential instability of an AdS small black hole, $\Lambda_{\text{BH}}$, has a domain of existence inside \begin{eqnarray}\label{rangeLambda}
\Lambda_{\rm min} < \Lambda_{\text{HP}} \lesssim \Lambda_{\text{BH}} \lesssim \Lambda_{\text{J}}\, . 
\end{eqnarray} 
In the limit $\left|\Lambda_{\text{AdS}} \right| \rightarrow 0$ (in $d$-dimensional Planck units), both energy scales vanish as 
\begin{eqnarray} 
\underbrace{\Lambda_{\rm min}(\phi) \simeq \Lambda_{\text{HP}}(\phi) \sim \left|\Lambda_{\text{AdS}}(\phi) \right|^{1/2}}_{\substack{\text{existence condition} \\ \rightarrow \text{min. temperature}}} \qquad  \underbrace{\Lambda_{\text{J}}(\phi) \sim \left|\Lambda_{\text{AdS}}(\phi) \right|^{1/d}}_{\substack{\text{gravitational collapse bound} \\ \rightarrow \text{max. temperature}}}, 
\end{eqnarray} 
which implies the following hierarchy of scales\footnote{To be precise, the bounds allow for a more general expression including multiplicative corrections
\begin{eqn}
    \Lambda_{\text{BH}} \sim \left|\Lambda_{\text{AdS}}\right|^\alpha \left( \left| \log^{\beta} \Lambda_{\text{AdS}} \right| + c + \sum_{n>0} a_n \Lambda_{\text{AdS}}^n \right)
\end{eqn}
where $c \neq 0$, and the exponent $\beta$ is constrained such that \eqref{rangeLambda} is satisfied; for instance if $\alpha = 1/2$ then $\beta \geq 0$, which agrees with the results obtained in \cite{Blumenhagen:2019vgj}.} 
\begin{equation}\label{ADCBH} 
\Lambda_{\text{BH}} \sim \left|\Lambda_{\text{AdS}}\right|^\alpha, \qquad \frac{1}{d} \leq \alpha \leq \frac{1}{2}\, .
\end{equation} 

As motivated in the previous chapters, it is now evident that the nature of (AdS-) black holes plays a crucial role in this version of the AdS Distance Conjecture. Specifically, their phase transitions provide a natural IR cutoff for $\Lambda_{\text{BH}}$, with the Hawking-Page transition---or the minimal temperature $\Lambda_{\rm min} = \mathcal{O}(1)\ \Lambda_{\text{HP}}\, $---serving as the lower bound, whereas the maximum EFT cutoff is given by $\Lambda_{\rm J}$. As first suggested in \cite{Bedroya:2024uva}, assuming a connection between the instability of neutral black holes and the emergence of a light tower of states in the weak gravity limit of the theory, the AdS Distance Conjecture naturally follows.
From this perspective, the latter establishes the existence of black hole instabilities in AdS space, and in the infinite-distance limit $\Lambda_{\text{AdS}} \rightarrow 0$, we observe universal bounds for the scaling behaviour of the scale of such instabilities. 
Moreover, as proposed in \cite{Bedroya:2024uva}, $\Lambda_{\text{BH}}$ can be viewed as a black-hole-corrected version of $m_{\rm tow}$, that reduces to the mass scale of towers in the known infinite distance limits in flat space.

As mentioned in section \ref{Introduction}, our argument relies solely on the assumption that the presence of a dynamical instability for neutral black holes in the gravitational weak gravitational-coupling limit is connected to the existence of an infinite tower of states, and viceversa. In particular, given that the different end-points of these instabilities generally depend on the specifics of the UV completion, we focused only on the existence of the instability. On top of that, the presence of a black hole interpretation for the mass scale of the tower, $m_{\text{tow}}$, allows to impose constraints on the existence of the latter. Hence, by not explicitly referencing the endpoints of such instabilities of black holes, we determined the general range of allowed values for the mass scales of the towers.\footnote{As argued above, this argument can be applied for generic solutions with a \textit{finite} non extremal-radius, i.e., non-extremal black hole.}

Following this logic, the scale of these instabilities is thus fixed in the weak gravity regime, $\Lambda_{\text{BH}} \sim m_{\text{tow}}$, by the following relation (in $d$-dimensional Planck units): 
\begin{equation} 
m_{\rm tow} \sim \left|\Lambda_{\text{AdS}}\right|^\alpha\, , \qquad \frac{1}{d} \leq \alpha \leq \frac{1}{2}\, . 
%M_{\text{Pl}, d}^{2\alpha -1}\, , 
\end{equation} 
Therefore, as discussed above, assuming the existence of the tower of states at $\Lambda_{\rm BH}$, the asymptotic behaviour of $m_{\rm tow}$ is constrained by black hole physics, as expected from the Anti-de Sitter Distance Conjecture. 

This approach does not explicitly depend on the notion of distance in moduli space, but instead focuses on the connection between the energy scales associated with the conjecture, allowing us to interpret these phenomena in terms of instabilities or phase transitions. Furthermore, if we include the emergent string conjecture \cite{Lee:2019wij}, which ought to apply in the flat space limit, the nature of these towers is fixed to be either a (dual of a) Kaluza-Klein tower or a tower of weakly-coupled string oscillators, and the instabilities are expected to correspond to the Gregory-Laflamme or the Horowitz-Polchinski, respectively. Conversely, if any dynamical instability was associated to a tower of states in the gravitational weakly coupled limit---such as the ones mentioned above---then it would be considered as the candidate to probe the corresponding tower of light states. The emergent string conjecture would then imply that if any other instabilities appears, then the Gregory-Laflamme and/or Horowitz-Polchinski ones ought to take place at a lower energy scale, possibly preventing the former to actually take place. 
However, our argument does not rely on the specifics of these two instabilities (or towers), and in principle remains valid for any other types of instabilities, possibly associated with different  towers of states.

\subsubsection*{Bound on the Number of Spacetime Dimensions}
\label{ss:MaxDim}
Additionally, using black hole arguments in the context of Swampland conjectures may offer further insights into their regime of validity. In pure AdS$_d$ space, different dynamical instabilities may not present for every dimension. Hence, one can bound the number of spacetime dimensions, $d$, from the requirement that they exist in the EFT, i.e., such that our argument in terms of black hole instabilities can work. Specifically, both the Jeans and Gregory-Laflamme/Horowitz-Polchinski instabilities may not occur in arbitrary dimensions.
In this section, we briefly recall some key results from the literature that allow us to establish such bounds on 
$d$, and we refer interested reader to \cite{Li:2009jq, Vaganov:2007at, Chavanis:2007kn ,Hammersley:2007ahw, Li:2008xw} for detailed information on the existence of such instabilities.

Indeed, in \cite{Li:2009jq, Vaganov:2007at, Chavanis:2007kn ,Hammersley:2007ahw, Li:2008xw} it was explicitly shown that the Jeans instability occurs for $4 \leq d \leq 10$. In this range, for instance in the canonical ensemble, every configuration with $T \gtrsim T_{\rm J}$ is unstable to collapse to a large black hole configuration.
However, for self-gravitating radiation in $d \geq 11$, the maximum temperature and mass, which would correspond to the Jeans temperature and the Jeans mass, can only be reached for singular energy densities---i.e., for infinite energy densities at the origin. Beyond this critical dimension, the instability is lost, and the bound on the temperature range, \eqref{rangeLambda}, no longer applies. Specifically, the Jeans instability occurs when $3 \leq d \leq 10$, and the thermal ensemble is thus unstable above $T_{\rm J}$.

The argument proposed in the previous section is built upon the existence of a bounded range of temperatures where black holes can become unstable, meaning we are concerned with the regime where the system is \textit{unstable}. This means that the upper bound for the black hole scale is dimension-dependent, and can only be directly applied in AdS$_d$ with $d < 11$. Therefore, the bound directly \eqref{ADCBH} holds  for: \begin{eqn} 3 \leq d \leq 10 , \end{eqn} implying the maximum range: \begin{eqn} \frac{1}{10} \leq \alpha \leq \frac{1}{2}. \end{eqn}

As a final remark, let us emphasize that similar dependence on $d$ can also be found for other kinds of instabilities \cite{Kol:2006vu, Sorkin:2004qq, Kol:2002xz, Harmark:2007md}, like those reviewed in section \ref{thermoInstabilitiesChapter}, which include the specific information about the towers that become light. In particular, using a Morse-theoretic approach, it was shown that not all black holes are unstable via a black hole-(generalized) black string transition, due to the different topologies of the horizons. 
For flat spacetimes with one compact dimension, the instability was shown to be present only for $d\leq 10$ \cite{Kol:2002xz}. Still, one should keep in mind that this last result is, in general, model-dependent and it may change with the introduction of new ingredients, different compactifications, or alternative transitions. However, when considering instabilities in the small black hole branch of AdS, these arguments are not expected to be significantly affected by the AdS geometry---especially in the small $\Lambda_{\rm AdS} \ll 1$ limit. Therefore, even though they must be taken with a grain of salt, some flat space considerations, like the ones presented above, may still apply. Nevertheless, studying those transitions for different compactifications may open a new window in view of this connection with the Swampland program.

\subsubsection*{Black hole perspective on scale separation}
Finally, let us point out that this relations also provide a neat perspective on the issue of scale separation in AdS vacua of the form AdS$_d \times \mathcal{Y}_p$, with $\mathcal{Y}_p$ a $p$-dimensional compact manifold, from the black hole perspective. From the discussion in section \ref{sss:GLandJeans} it follows that an AdS vacuum is scale-separated if a large AdS black hole becomes thermodynamically favourable at the same (parametric) scale  at which a small black hole suffers from the Gregory-Laflamme instability. In other words, there is no scale separation between the two instabilities: 
\begin{equation} 
\frac{\Lambda_{\text{HP}}(\phi)}{\Lambda_{\text{BH}}(\phi)} \sim 1. 
\end{equation}
This implies the impossibility of describing any small black hole in supergravity at energy scales above  $\Lambda_{\text{BH}}$, since the details of the dynamics after the instability require information about its profile along the extra dimensions, $\mathcal{Y}_p$, which cannot be described in the AdS$_d$ EFT (gauged supergravity). The AdS Distance Conjecture thus indicates how the two instabilities are interconnected in that particular limit. 

If $\alpha > 1/2$, the Gregory-Laflamme instability of the small black hole would occur at an energy scale below that associated to the presence of the first small black hole. This would imply a drastic change in the phase transitions of  \bh, which directly prevents us from using a lower EFT description.\footnote{As discussed in previous sections, for non-scale separated models, such as AdS$_5 \times S^5$, it has been argued in \ref{thermoInstabilitiesChapter} that the Gregory-Laflamme instability happens slightly \textit{after} the Hawking-Page transition, by just a $\mathcal{O}(1)$ factor.}

On the contrary, a vacuum is scale-separated if 
\begin{equation} \frac{\Lambda_{\text{HP}}(\phi)}{\Lambda_{\text{BH}}(\phi)} \ll 1\, . 
\end{equation}
which implies $\alpha < 1/2$. Moreover, the largest allowed range of scale separation from the requirement of existence of small black holes is identified with
\begin{equation} 
\frac{\Lambda_{\text{J}}(\phi)}{\Lambda_{\text{BH}}(\phi)} \sim 1, 
\end{equation} 
where $\Lambda_{\text{J}}(\phi)$ is the (possibly quantum-corrected) moduli-dependent Jeans temperature, related to the bound $\alpha \geq 1/d$.

\subsection{Black Holes and the AdS Distance Conjecture in the Interior of Moduli Space}\label{alphaprimecorr}
\label{ss:higherderivative}

As discussed in the previous sections, our argument relates the domain of validity of $\Lambda_{\text{BH}}$ in AdS backgrounds to the  AdS distance conjecture. In the context of black holes, it is natural to study how the scales of instability change in the presence of higher-derivative corrections within the EFT. Hence, this gives us a handle to understand how the upper and lower bounds obtained from our approach---and consequently the AdS Distance Conjecture---get modified as we move away from the asymptotic regimes---i.e. in the interior of moduli/field space.

To this end, we study how the Hawking-Page transition (and the corresponding critical temperature) gets modified in the presence of (higher derivative-)corrections. Our analysis here is simply meant to provide a proof of concept, as opposed to a general and systematic analysis, so to this end we consider the particular example of $\mathcal{O}(\alpha^{\prime 3})$ corrections to the AdS$_5$ black hole entropy stemming from the $R^4$ term in the effective action of type IIB string theory \cite{Gubser:1998nz, Caldarelli:1999ar, Landsteiner:1999gb, Dutta:2006vs}. In the Einstein frame, using the democratic formalism, the type IIB string effective action in 10d has the following structure:
\begin{eqn}
    S = {\frac{1}{16 \pi G_{10}}} \int d^{10}x\, \sqrt{-g_{10}} \left(R - \frac{1}{2} (\partial \Phi)^2
- {\frac{1}{4 \cdot 5!}} (F_5)^2 + \gamma e^{-\frac{3}{2}\Phi} W + \dots \right)
\end{eqn}
where  $ \gamma = \frac{1}{8} \zeta(3) \alpha^{\prime 3}$, and $W$ denotes the $R^4$ term
\begin{eqn}\label{eq:W}
    W = R^{abcd}R_{ebcf}R^{hie}_{a}R^{f}_{hid} +  \frac{1}{2}R^{abcd}R_{efab}R^{ghe}_{a}R^{b}_{ghf} + \text{Ricci tensor part}.
\end{eqn}
Field reparametrization invariance allows one to change the coefficients of the terms involving the Ricci tensor and rewrite $W$ in terms of the Weyl tensor.\footnote{It has been argued that possible extra $F_5$-dependent terms should not affect the computation since the 5-form for the non-extremal background is the same as when the non-extremal radius is null, $r_0 = 0$.\cite{Gubser:1998nz}}
Then, we can write  the AdS$_5 \times S^5$ background, but recalling that in presence of higher derivative corrections one needs to use the perturbed metric ansatz 
\begin{eqn}
    ds^2 = H^2(r)\left( K^2(r)dt^2 + P^2(r) dr^2 + \ell_{\rm AdS}^2 d\Sigma^2 \right)\,  .
\end{eqn}
This allows one to express the corrected action as
\begin{eqn}
    S = -\frac{1}{16\pi G_5} \int_{r_0}^\infty dr \sqrt{-g_5}\left[\underbrace{ \;\; R + \frac{12}{\ell_{\rm AdS}^2} \;\; }_{\text{tree-level + $F_5$-flux}} - \underbrace{\frac{1}{2}(\partial_r \phi(r))^2 + \gamma e^{-\frac{3}{2}\phi} W + \dots}_{\text{corrections}}\right] \, ,
\end{eqn}
where ${\ell_{\rm AdS}^4} = (2g_s N){\alpha^{\prime 2}}$,  $G_5 \ell_{\rm AdS}^5 = G_{10}$, $N$ labels the units of $F_5$ flux and we can define the large parameter $\lambda=\sqrt{g_s \, N}$.
With the perturbed metric obtained in \cite{Gubser:1998nz, Caldarelli:1999ar}, one can then  introduce the requirement that it does not have a conical singularity at the horizon, $r_+$. In this way, the corrected period $\beta = T^{-1}$ reads
\begin{equation}
    T = {2 r_+^2 + 1\over 2\pi r_+} \left( 1 + \gamma {10 (r_+^3+1)^3
(3r_+^2-1) \over r_+^6 (2r_+^2 + 1)} \right).
\end{equation}
This yields the following expression for the free energy 
\begin{eqn}\label{freeEnergyCorr}
    F =
{\mathcal{V}_{S^3} \mathcal{V}_{S^5} \over 16\pi G_{10}} \left[ r_+^4 - r_+^2 + 5 \gamma
{(r_+^2+1)^3 (15r_+^2 -1) \over r_+^4} \right]\,.
\end{eqn} 
The critical black hole radius that follows from this expression is ${r_+^c = (1 - 280 \gamma) \ell_{\rm AdS}}$, and the corresponding minimum temperature is
\begin{eqn}
    \Lambda_{\rm min } = \frac{9}{4 \sqrt{2}\pi \ell_{\rm AdS}} \left(1 - 5 \pi  \zeta(3) \frac{\alpha^{\prime 3}}{\ell_{\rm AdS}^6} \right) + \mathcal{O}(\lambda^{-6}).
\end{eqn}
Finally, the one-loop corrected Hawking-Page temperature is 
\begin{eqn}
     \Lambda_{\rm HP} = {3\over 2\pi \ell_{\rm AdS}} \left(1 - \frac{1}{5} \zeta(3)  \frac{\alpha^{\prime 3}}{\ell_{\rm AdS}^6} \right) + \mathcal{O}(\lambda^{-6}).
\end{eqn}

Recall that  ${\ell_{\rm AdS}}\,  M_{\rm s} \sim \sqrt\lambda = (g_s N)^{1/4}$, and additionally $\ell_{\rm AdS} \, M_{\mathrm{Pl,}\,5 }\sim N^{2/3}$.
The limit $\Lambda_{\rm AdS}/ M_{\mathrm{Pl,}\,5 }^2\ll 1$ can thus be achieved for any $N\gg 1$, and if also $\lambda \gg 1$, one has a decompactification to 10d flat space, in which $ \ell_{\rm AdS}^{-1} \lesssim M_{\rm s}\lesssim M_{\mathrm{Pl,}\, 10}$. In our AdS$_5\times S^5$ setup, where there is no scale-separation, $m_{\rm KK} \sim  \ell_{\rm AdS}^{-1}$, and in the $\lambda \gg1$ limit one obtains
\begin{equation}\label{HPcorrections}
    \frac{\Lambda_{\rm HP}}{m_{\rm KK}} \propto \left( 1 - \frac{\zeta(3)}{5\lambda^3}\right) \, , 
\end{equation}
where the quantity inside the parenthesis is smaller than one. Using the results from section \ref{sss:GLandJeans}, at tree-level $\Lambda_{\rm BH} > \Lambda_{\rm HP}$. Hence, due to the fact that the corrections only lower the ratio $\Lambda_{\rm HP}/m_{\rm KK}$, it follows that the ordering $m_{\rm KK} \sim \Lambda_{\rm BH}>\Lambda_{\rm HP}$ is still maintained.

Similar arguments can be applied for M-theory on AdS$_4 \times S^7$, and AdS$_7 \times S^4$, considering higher-derivative terms for the near-horizon limit of M$_2$ and M$_5$ branes. In this case, the corrections will depend on $M_{{\rm Pl}, 11}$ \cite{Gubser:1998nz, Landsteiner:1999gb, Caldarelli:1999ar}

Then, at fixed cosmological constant, in this case supported by the $F_5$-flux, $N$, the black hole-related scale, $T_{\rm HP}$, acquires corrections from the higher-curvature term. These corrections are negative, and the Hawking-Page, as well as the minimal temperature, stay below the bare $\Lambda_{\rm AdS}^{1/2}$. Hence, when corrections are added, it is possible to produce black holes in AdS at smaller energies. 
As shown in \cite{Bedroya:2024uva}, the black-hole quantity $\Lambda_{\rm BH}$ differs from $m_{\rm KK}$ in the presence of higher derivative corrections, possibly interpolating between the Gregory-Laflamme instability and the Horowitz-Polchinski one. Similarly, the ratio between the Hawking-Page (or the minimal temperature) and the AdS cosmological constant is also corrected, such that $\Lambda_{\rm HP}/\Lambda_{\rm AdS}^{1/2} \sim \left[1 - \mathcal{O}(\lambda^{-3})\right]$. The Hawking-Page temperature is thus affected by the higher curvature corrections and reduces to t $\Lambda_{\rm AdS}^{1/2}$ in the small cosmological constant limit. The key insight is then that this temperature gives a quantity that we can use to promote $\Lambda_{\rm AdS}^{1/2}$ to a phase transition scale for black holes, given by $T_{\rm HP}(\phi)$, which can be defined not only asymptotically. In this way, this scale may represent a moduli-corrected lower bound to $\Lambda_{\rm BH} (\phi)$ (or $\Lambda_{c}(\phi)$), that reduces to  $\Lambda_{\rm AdS}^{1/2} \lesssim m_{\rm KK}$ asymptotically. 
Moreover, higher-derivative corrections are expected to affect $T_{\rm J}$. For small corrections, the saddle point \eqref{eq:tempCanonical} will be modified because of the different metric, which change the local temperature due to corrections to $V(r)$ in e.g. \eqref{eq:emptyAdSmetric}. From dimensional analysis we could expect a correction of the form $T_{\rm J}^{\text{corr.}}\simeq T_{\rm J}(1 -\mathcal{O}\left(\ell/{\ell_{\rm AdS}})^{n} \right)$ with $n \geq 2$, with $\ell$ some cutoff scale suppressing the relevant higher curvature operators. We leave this calculation and its physical interpretation for future work.

 \subsection{On the Cosmological Horizon}
\label{ss:cosmologicalhorizons}

de Sitter (dS) spacetime is the well-known maximally symmetric solution to Einstein’s equations with a positive cosmological constant. Additionally, it has a causally non-trivial spacetime structure with observer-dependent horizons. For instance in de Sitter spacetime, the location of the horizon is not physically unique, meaning that the question of the stability of the cosmological horizon is more subtle \cite{Hubeny:2002xn}.
However, given that cosmological horizons share some similarities with black hole horizons, one may expect that the de Sitter horizon could also exhibit dynamic instabilities for a given observer, much like its black hole counterparts.

Building on these considerations, it is natural to explore whether metric perturbations related to extra dimensions could also induce instabilities while preserving de Sitter symmetry. This would require altering the size of the de Sitter spacetime. Such a scenario can be viewed as analogous to spherically symmetric perturbations in dimensionally reduced black holes, where the horizon size becomes non-uniform in the transverse directions \cite{Hubeny:2002xn, Bousso:2002fi}. 
For example, as shown by these authors, a spacetime of the form $dS_p \times S^q$ is perturbatively unstable with respect to the presence of $S^q$ due to a Gregory-Laflamme instability.\footnote{This setup could be interpreted as representing only a portion of the total number of compact dimensions, which could in a general be accompanied by a negatively curved manifold.} Furthermore, the configuration $dS_p \times S^q$ is not be the most entropically favourable. From a global entropy perspective, one might expect that $dS_p \times S^q$ would tend to localize into pure $dS_{p+q}$ \cite{Hubeny:2002xn}. While this may seem counter-intuitive, it is similar to what occurs in black hole systems. The key distinction lies in the fact that for a $(d+p)$-dimensional black hole, the configuration becomes more entropically favourable when $r_h \simeq R$ (with $R$ the $S^p$ radius) potentially localizing into the extra dimensions. Conversely, in the case of de Sitter, the extra dimensions may be included when $\ell_{\rm dS} \simeq R$\cite{Hubeny:2002xn}.

One can also investigate spacetimes that are only asymptotically de Sitter. The next-to simplest case is a Schwarzschild-de Sitter spacetime. In $d$-dimension, the SdS metric in static coordinates is given by
  \begin{align}
  ds^2 &= -f(r) dt^2 + f^{-1}(r)dr^2 + r^2 d \Omega_{d-2}^2 \, , \quad  \label{metric1} \\ 
  f (r) &= 1 - \frac{r^2}{\ell_{\rm dS}^2} - \frac{G_N M  }{ r^{d-3}}\,. \label{eq:blackeningfactor}
  \end{align}
Here,  $\ell_{\rm dS}=\sqrt{( d-1)(d-2) /(2 \Lambda_{\rm dS})}$ is the de Sitter curvature radius. Requiring the absence of a naked singularity yields an upper limit on the mass of SdS black holes
\begin{equation} \label{eq:range}
    0 \le M \le M_{\rm N} \,, \qquad \text{with} \quad M_{\rm N} =  \frac{ \alpha_d}{8\pi G_N} \ell_{\rm dS}^{d-3} \,,
\end{equation}
where $\alpha_d = \frac{d-2}{d-1}  \left(\sqrt{\frac{d-3}{d-1}}\right)^{{d-3}} $ depends only on the number of dimensions $d$. The case $M=0$ corresponds to pure de Sitter space, which has a single (cosmological) event horizon located at $r_{h, \mathrm{dS}}=\ell_{\rm dS}$.
For the values  $0< M < M_{\rm N}$ the function $f(r)$ has two positive  real roots $r_b$ and $r_c$, with $r_b \le r_c$, corresponding to the position of the black hole event horizon  and the cosmological event horizon.  As $M$ increases, the black hole horizon increases in size, whereas the cosmological horizon shrinks in size due to the gravitational pull of the black hole. The  upper bound $M=M_N$ corresponds to the case where the black hole and cosmological horizon coincide, $r_b=r_c=r_{\rm N}$, known as the Nariai solution \cite{Nariai:1999iok}.
In the context of perturbative spacetime instabilities, is conceivable that if a black hole is dynamically unstable and its event horizon approaches or even coincides with the cosmological horizon, the cosmological horizon could similarly experience a dynamical instability. Otherwise, the growth of the black hole's horizon area without a corresponding increase in the de Sitter horizon area might potentially violate the N-holographic bound \cite{Bousso:2022ntt}. This suggests that the temperature of the Nariai black hole $T_{\rm N}$ \cite{Bousso:1997wi} would be a lower bound for $\Lambda_{\rm BH}$
\begin{eqn} 
T_{\rm N} \sim \Lambda_{\rm dS}^{\frac{1}{2}} \lesssim \Lambda_{\rm BH}.
\end{eqn} 
As discussed in the previous sections, assuming that the dynamical instability is associated with a tower of states in the $\Lambda_{\rm dS} \to 0$ limit, we would trivially arrive at a bound similar to the Higuchi bound \cite{Higuchi:1986py}: \begin{eqn} \Lambda_{\rm BH} \sim m_{\rm tow} \gtrsim \Lambda_{\rm dS}^{\frac{1}{2}}. \end{eqn}
Generally speaking, neutral black hole instabilities stem from the presence of a horizon and are tied to its thermodynamic properties. For a universe that is asymptotically de Sitter, it is reasonable to extend the scale of instability for a neutral black hole to a more generic horizon instability.
So, it would be natural to expect to apply similar bounds to those in previous sections for the de Sitter case. 

However, the same argument of section \ref{ADC} cannot be directly applied in this context. Indeed, thermal dS is stable against gravitational collapse of thermal radiation, since at fixed temperatures for configurations in which thermal dS dominates, the finite size of the cavity is smaller than the Jeans length, preventing the collapse. Indeed, it has been showed that for classical perturbations around the de Sitter metric there is no analogue of the Jeans instability \cite{Ginsparg:1982rs}. Roughly speaking, due to the acceleration, the rate of collapse is always less than the rate of expansion. However, it is not clear if any instability may  appear at the non-linear level. 
This would imply that only in Anti-de Sitter spacetime we have a well-defined thermal ensemble with such instability, which is intrinsically connected to the BF bound.\footnote{Differently from AdS, for (semi-classical)dS, due to the finiteness of its entropy, there is another instability which is due to the spontaneous nucleation of black holes pairs\cite{Ginsparg:1982rs} arguing that this equilibrium would be unstable.}

For the reasons expressed above, a natural set-up with positive vacuum energy which possibly avoids the absence of the Jeans instability may be a non-accelerating universe. Hence, the only cosmology such that a Jeans-like instability may be present would correspond to a universe with scale factor of the form $a(t) = t^p$, with $p < 1$. For instance, if we assume that the scalar fields are the main driver of the expansion, then the latter case would correspond to a scalar potential driven by some fields $\phi_i$, giving $V(\phi_i) \, = \, V_0\ e^{-\lambda \phi}$. Surprisingly, as argued in \cite{Bedroya:2022tbh}, the case $p<1$ corresponds to $\lambda \geq \frac{2}{d-2}$, recovering the TCC bound\cite{Bedroya:2019snp}. It would be extremely interesting to explore these connections further.

\section{Conclusions and Outlook}
\label{s:Summary}

The objectives of this work are threefold: (i) To reinforce the perspective that black hole instabilities serve as effective probes for exploring Swampland constraints and vice versa, providing an explicit bottom-up derivation of the AdS Distance Conjecture. (ii) To establish that horizon instabilities, particularly black hole instabilities, act as powerful diagnostics for key scales within effective field theories (EFTs), such as the Kaluza–Klein scale, $m_{\rm KK}$, and the string scale, $M_{\rm s}$, along the lines of \cite{Bedroya:2024uva}. (iii) To introduce a moduli-corrected extension of the scales that bounds $\Lambda_{\rm BH}$ in AdS backgrounds by accounting for higher-derivative corrections within the gravitational sector of EFTs, allowing for a formulation of such bounds in the interior of moduli space.

Firstly, black holes, as intrinsically quantum-gravitational systems, offer an ideal setting for systematically studying the physics associated with UV/IR mixing within semiclassical gravity. Specifically, we focused on Schwarzschild-AdS black holes in $d$ dimensions due to the rich and controlled framework provided by the AdS background, as well as the inherent constraints highlighted by the classical Hawking–Page analysis \cite{Hawking:1982dh}. In particular, we used two key insights from the latter analysis: the existence of Schwarzschild-AdS$_d$ black holes only above a certain minimal temperature, $T_{\rm min}\simeq T_{\rm HP}\sim |\Lambda_{\rm AdS}|^{1/2}$, and the fact that small black holes in equilibrium with thermal radiation become unstable and collapse into large black holes above the Jeans temperature, $T_{\rm J}\sim |\Lambda_{\rm AdS}|^{1/d}$.

Secondly, through detailed examination of small black hole and  large black hole solutions in equilibrium with thermal radiation in AdS, both in the canonical and microcanonical ensembles, we derived robust and universal relationships among various instability scales. Our findings align with recent proposals \cite{Bedroya:2024uva}, suggesting that neutral black hole instabilities at scales ${\Lambda_{\rm BH} \ll 1}$ (in Planck units) indicate the emergence of a tower of states, with the lowest mass scale given by $m_{\rm tow} \sim \Lambda_{\rm BH}$. Consequently, we obtained the refined version of the Anti-de Sitter Distance Conjecture:
\begin{equation}
    \Lambda_{\rm BH}\sim |{\Lambda_{\rm AdS}}|^{\alpha}\,,   \qquad {\rm with }\qquad \frac{1}{d}\leq \alpha\leq \frac{1}{2}\, ,
\end{equation}
as $\Lambda_{\rm AdS}\to 0$ in Planck units, purely in terms of physical scales related to black holes and the AdS background, rather than relying explicitly on relationships involving $\Lambda_{\rm AdS}$ and field-space distances, $\Delta_\phi$. Additionally, we emphasized that our formulation of the AdS Distance Conjecture focuses explicitly on the existence of an instability without specifying its final state. This formulation avoids subtleties related to the endpoints of black hole evolution, thus providing universal bounds for the scale  $\Lambda_{\rm BH}\sim m_{\rm tow}$ in asymptotic limits.

Furthermore, from a top-down perspective the transition that takes place at the scale $\Lambda_{\rm BH}$ can be argued to correspond to either the Gregory-Laflamme or Horowitz-Polchinski transition in the flat space case \cite{Bedroya:2024uva}. We found that for asymptotically AdS backgrounds, the same behaviour is recovered in the limit $\Lambda_{\rm AdS}\to 0$, where the black hole scale reduces to that of the tower and is still comprised in the region between the Hawking-Page temperature and the Jeans temperature.

Thirdly, we analyzed known $\alpha^\prime$-corrections within the explicit context of type IIB string theory on AdS$_5\times S^5$ \cite{Gubser:1998nz, Caldarelli:1999ar, Landsteiner:1999gb, Dutta:2006vs}. This serves as a proof of concept that higher-derivative corrections modify the bounds on $\Lambda_{\rm BH}$, an essential step for interpreting these constraints as we move towards the interior of moduli space. Our analysis enriches the understanding of scaling behaviors characterized by the exponent $\alpha$ in various instabilities and phase transitions. Importantly, we highlighted the validity of the conventional identification $m_{\rm tow} \sim \Lambda_{\rm AdS}^{\alpha}$ in the strict limit $\Lambda_{\rm AdS} \to 0$, at least as long as $M_{\rm s}\gg |\Lambda_{\rm AdS}|^{1/2}$. It would be interesting to perform a more detailed and systematic analysis of these cases, also in the light of the recent results for tensionless string limits \cite{Calderon-Infante:2024oed}, where the condition $M_{\rm s}\gg |\Lambda_{\rm AdS}|^{1/2}$ can be relaxed, in the context of the CFT Distance Conjecture \cite{Baume:2020dqd,Perlmutter:2020buo,Baume:2023msm}.

An unexpected implication of our analysis is the potential breakdown of the Anti-de Sitter Distance Conjecture for dimensions $d>10$, arising from distinct behaviors of Schwarzschild-AdS$_d$ solutions in higher dimensions, studied in \cite{Li:2009jq, Vaganov:2007at, Chavanis:2007kn ,Hammersley:2007ahw, Li:2008xw}. Conversely, it is tantalizing to promote this observation to an indirect upper bound on the number of non-compact dimensions, $d\leq 10$, and further understanding the implications of this reasoning is left for future work.

Finally, we briefly extended some aspects of our discussion to include an expanding universe---including the de Sitter case---underlining the similarities and differences relative to the Anti-de Sitter case. We outlined the broader implications of our results for the Swampland program, emphasizing how insights from black hole instabilities may guide future explorations of conjectures across diverse spacetime backgrounds. A detailed investigation of the implications of these insights, particularly regarding the behavior of instabilities and associated constraints in de Sitter space, remains an open and interesting avenue to pursue.

\vspace{1.5cm}
\centerline{\bf Acknowledgments}
\vspace{-0.2cm}
We are grateful to I.~Basile, A.~Castellano, E.~Kiritsis, J.~Masias, C.~Vafa and M.~Zatti for useful discussions. The work of D.L. is supported by the Origins Excellence Cluster and by the German-Israel-Project (DIP) on Holography and the Swampland.

\newpage
\bibliography{references} 
\bibliographystyle{JHEP}
\end{document}